\def\figsize{9.5cm}

\def\rn{}
\def\nn#1 #2{#2. #1}				
\def\nnn#1 #2 #3{#2. #3. #1}			
\def\nnnn#1 #2 #3 #4{#2. #3. #4 #1}		
\def\nnnnn#1 #2 #3 #4 #5{#2. #3. #4 #5. #1}	
\def\dualand{ and\hbox{ }}				
\def\multiand{, and\hbox{ }}				
\def\rf#1;#2;#3;#4;#5 {{\frenchspacing\par\rn#1, #3 {\bf #4}, #5 (#2). \par}}
\def\rg#1;#2;#3;#4;#5;#6 {{\frenchspacing\par\rn#1, #3 {\bf #4}, #5 (#2). \par}}
\def\rfbook#1;#2;#3;#4;#5 {{\frenchspacing\par\rn#1, {\it #3} (#5, #4, #2).\par}}
\def\rfprep#1;#2;#3 {{\par\frenchspacing\rn#1, #3 (#2).\par}}
\def\rfproc#1;#2;#3;#4;#5;#6 {{\frenchspacing\par\rn#1 #2, in {\it #3}, ed. #4 (#5: #6)\par}}
\def\rfprocp#1;#2;#3;#4;#5;#6;#7 {{\frenchspacing\par\rn#1 #2, in {\it #3}, ed. #4 (#5: #6), p#7\par}}

\def\rg#1;#2;#3;#4;#5;#6 {\par\rn#1 #2, {\it #3}, {\bf #4}, #5 (``#6'') \par}
\def\rf#1;#2;#3;#4;#5 {\par\rn#1, {\it #3}, {\bf #4}, #5 (#2)\par}
\def\rfbook#1;#2;#3;#4;#5 {{\frenchspacing\par\rn#1, {\it #3} (#4: #5, #2)\par}}
\def\rfproc#1;#2;#3;#4;#5;#6 {{\frenchspacing\par\rn#1 #2, in {\it #3}, ed. #4 (#5: #6)\par}}
\def\rfprocp#1;#2;#3;#4;#5;#6;#7 {{\frenchspacing\par\rn#1 #2, in {\it #3}, ed. #4 (#5: #6), p#7\par}}
\def\rfprep#1;#2;#3  {{\par\rn#1, #3, #2\par}}
\def\rfprepp#1;#2;#3 {{\par\rn#1 #2, #3\par}}

\def\kg{{\rm kg}}
\def\Meter{{\rm m}}
\def\Second{{\rm s}}
\def\Gyr{{\rm Gyr}}
\def\K{{\rm K}}

\def\mK{{\rm \mu K}}

\def\pc{{\rm pc}}
\def\Mpc{{\rm Mpc}}

\def\km{{\rm km}}

\def\eV{{\rm eV}}

\def\MeV{\>{\rm MeV}}
\def\GeV{\>{\rm GeV}}

\def\expec#1{\langle#1\rangle}

\def\etal{{\frenchspacing\it et al.}}
\def\ie{{\frenchspacing\it i.e.}}
\def\eg{{\frenchspacing\it e.g.}}
\def\etc{{\frenchspacing\it etc.}}
\def\rms{{\frenchspacing r.m.s.}}

\def\beq#1{\begin{equation}\label{#1}}
\def\eeq{\end{equation}}
\def\beqa#1{\begin{eqnarray}\label{#1}}
\def\eeqa{\end{eqnarray}}
\def\eq#1{equation~(\ref{#1})}
\def\Eq#1{Equation~(\ref{#1})}
\def\eqn#1{~(\ref{#1})}

\def\fig#1{Figure~\ref{#1}}
\def\Fig#1{Figure~\ref{#1}}

\def\FundParTable{1}
\def\DerivedParTable{2}
\def\DerivedQuantTable{3}
\def\ConstraintTable{4}

\def\Sec#1{Section~\ref{#1}}
\def\App#1{Appendix~\ref{#1}}

\def\spose#1{\hbox to 0pt{#1\hss}}
\def\simlt{\mathrel{\spose{\lower 3pt\hbox{$\mathchar"218$}}
     \raise 2.0pt\hbox{$\mathchar"13C$}}}
\def\simgt{\mathrel{\spose{\lower 3pt\hbox{$\mathchar"218$}}
     \raise 2.0pt\hbox{$\mathchar"13E$}}}
\def\simpropto{\mathrel{\spose{\lower 3pt\hbox{$\mathchar"218$}}
     \raise 2.0pt\hbox{$\propto$}}}

\def\ed{\end{document}}

\def\curvature{\kappa}
\def\Otot{\Omega_{\rm tot}}
\def\Ob{\Omega_{\rm b}}
\def\Oc{\Omega_{\rm cdm}}

\def\Ol{\Omega_\Lambda}
\def\Om{\Omega_{\rm m}}
\def\Od{\Omega_{\rm d}}
\def\On{\Omega_\nu}
\def\ob{\omega_{\rm b}}
\def\ocdm{\omega_{\rm cdm}}
\def\od{\omega_{\rm d}}
\def\og{\omega_\gamma}

\def\olam{\omega_\Lambda}
\def\om{\omega_{\rm m}}
\def\on{\omega_\nu}
\def\fb{f_{\rm b}}
\def\fn{f_\nu}
\def\ns{{n_{\rm s}}}
\def\nt{{n_{\rm t}}}
\def\al{\alpha_n}

\def\As{A_{\rm s}}

\def\Mnu{M_\nu}
\def\mp{m_{\rm p}}
\def\mn{m_{\rm n}}

\def\trd{t_{\rm RD}}
\def\tmd{t_{\rm MD}}
\def\tvd{t_{\rm \Lambda D}}

\def\alphaw{\alpha_{\rm w}}

\def\alphas{\alpha_{\rm s}}
\def\thetaqcd{\theta_{\rm qcd}}
\def\Lambdaqcd{\Lambda_{\rm qcd}}
\def\rhoplanck{\rho_{\rm planck}}

\def\Ry{\rm Ry}
\def\aB{a_{\rm B}}
\def\st{\sigma_{\rm t}}

\def\Rc{R_{\rm c}}

\def\Rn{R_\nu}

\def\rhob{\rho_{\rm b}}
\def\rhoc{\rho_{\rm c}}

\def\rhon{\rho_\nu}
\def\rhong{\rho_\nu^\gamma}
\def\rhol{\rho_\Lambda}
\def\rhog{\rho_\gamma}
\def\rhom{\rho_{\rm m}}
\def\rhok{\rho_{\rm k}}
\def\rhostar{\rho_*}
\def\rhostarbar{\hat\rho_*}
\def\rhoh{\rho_h}
\def\rhomax{\rho_{\rm max}}
\def\Hstar{H_*}

\def\xib{\xi_{\rm b}}
\def\xic{\xi_{\rm c}}
\def\xid{\xi_{\rm d}}
\def\xin{\xi_\nu}
\def\xiwimp{\xi_{\rm wimp}}
\def\xiaxion{\xi_{\rm axion}}
\def\fnu{f_\nu}
\def\mwimp{m_{\rm wimp}}
\def\nwimp{n_{\rm wimp}}
\def\rhowimp{\rho_{\rm wimp}}
\def\nb{n_{\rm b}}
\def\ng{n_\gamma}
\def\nnu{n_\nu}
\def\aeq{a_{\rm eq}}

\def\Avac{A_\Lambda}

\def\Teq{T_{\rm eq}}
\def\rhomeq{\rhom^{\rm eq}}
\def\xeq{x_{\rm eq}}

\def\kmin{\kappa_{\rm min}}
\def\Amax{A_{\rm max}}
\def\xturn{x_{\rm turn}}
\def\tturn{t_{\rm turn}}
\def\sigmastar{\sigma_*}
\def\Einfl{E_{\rm inf}}

\def\Teq{{T_{\rm eq}}}
\def\Gl{{G_\Lambda}}
\def\Ginf{{G_\infty}}

\def\mpl{m_{\rm pl}}
\def\Tmin{{T_{\rm min}}}
\def\Mmin{{M_{\rm min}}}
\def\Esn{{E_{\rm sn}}}
\def\fa{f_a}

\def\Ms{M_\odot}
\def\tento#1{\times 10^{#1}}

\def\beq#1{\begin{equation}\label{#1}}
\def\eeq{\end{equation}}
\def\beqa#1{\begin{eqnarray}\label{#1}}
\def\eeqa{\end{eqnarray}}
\def\eq#1{equation~(\ref{#1})}
\def\Eq#1{Equation~(\ref{#1})}
\def\eqn#1{~(\ref{#1})}

\def\p{{\bf p}}

\def\x{{\bf x}}

\def\J{{\bf J}}

\def\ignore#1{}

\def\Mmin{{M_{\rm min}}}

\def\rhoturn{\rho_{\rm turn}}
\def\rhovir{\rho_{\rm vir}}
\def\rhomin{\rho_{\rm min}}
\def\rhomax{\rho_{\rm max}}
\def\nvir{n_{\rm vir}}
\def\Tvir{T_{\rm vir}}
\def\vvir{v_{\rm vir}}

\def\fmx{f_{\mu x}}
\def\nstar{n_{\star}}
\def\Nstar{N_{\star}}
\def\nstarfudge{f_\star}
\def\Meq{M_{\rm eq}}
\def\Mbh{M_{\rm b}^{\rm h}}
\def\rau{r_{\rm au}}
\def\torb{t_{\rm orb}}
\def\fprior{f_{\rm prior}}
\def\fselec{f_{\rm selec}}

\def\Mdeath{M_\dagger}
\def\sigmadeath{\sigma_\dagger}
\def\vdeath{v_\dagger}
\def\deathrate{\gamma_\dagger}

\def\mstar{m_\star}

\newcommand{\erfc}{{\mathrm{erfc}}}

\def\simless{\mathbin{\lower 3pt\hbox
        {$\,\rlap{\raise 5pt\hbox{$\char'074$}}\mathchar"7218\,$}}} 
\def\simgreat{\mathbin{\lower 3pt\hbox
        {$\,\rlap{\raise 5pt\hbox{$\char'076$}}\mathchar"7218\,$}}} 

\documentclass[twocolumn,amsmath,nofootinbib]{revtex4} 
\usepackage{url}
\begin{document}
\input{epsf.sty}

\def\affilmrk#1{$^{#1}$}
\def\affilmk#1#2{$^{#1}$#2;}

\title{Dimensionless constants, cosmology and other dark matters}

\author{Max Tegmark$^{1,2}$, Anthony Aguirre$^3$, Martin J.~Rees$^4$ \& Frank Wilczek$^{2,1}$}
\address{$^1$ MIT Kavli Institute for Astrophysics and Space Research, Cambridge, MA 02139}
\address{$^2$ Dept. of Physics, Massachusetts Institute of Technology, Cambridge, MA 02139}
\address{$^3$Department of Physics, UC Santa Cruz, Santa Cruz, CA 95064}
\address{$^4$Institute  of Astronomy, University of Cambridge, Cambridge CB3 OHA, UK}

\date{Submitted December 1 2005; accepted December 7; Phys.~Rev.~D, {\bf 73}, 023505}

\begin{abstract}
We identify 31 
dimensionless physical constants required by particle physics and cosmology, and emphasize that both microphysical constraints and
selection effects might help elucidate their origin. Axion cosmology provides an instructive example, in which these two kinds of
arguments must both be taken into account, and work well together. 
If a Peccei-Quinn phase transition occurred before
or during inflation, then the axion dark matter density will vary from place to place with a probability distribution.
By calculating the net dark matter halo formation rate as a function of all four relevant cosmological parameters and assessing other constraints,
we find that this probability distribution, computed at stable solar systems, is arguably peaked near the observed dark matter density. 
If cosmologically relevant WIMP dark matter is discovered,
then one naturally expects comparable densities of WIMPs and axions, making it important 
to follow up with precision measurements to determine whether WIMPs account for all of the dark matter or merely part of it.
\end{abstract}


  
\maketitle

\def\xistar{\xi_*}
\setcounter{footnote}{0}

\begin{figure}[pbt]
\vskip-0.5cm
\centerline{{\vbox{\epsfxsize=9.0cm\epsfbox{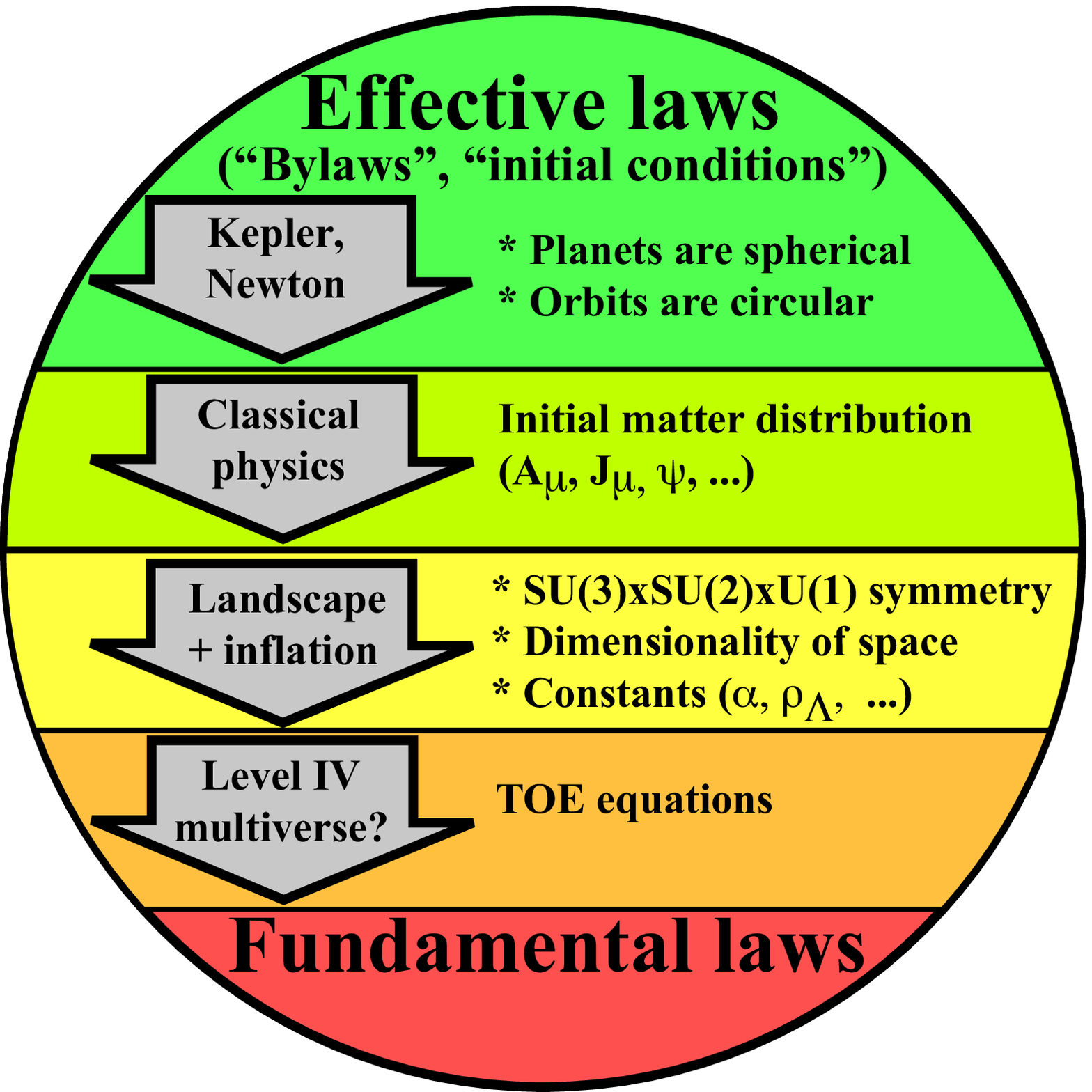}}}}
\vskip-0.3cm

\caption{\label{BoundaryFig}
The shifting boundary (horizontal lines) 
between fundamental laws and environmental laws/effective laws/initial conditions.
Whereas Ptolemy and others sought to explain roughly spherical planets and circular orbits
as fundamental laws of nature, Kepler and Newton reclassified such properties as initial conditions
which we now understand as a combination of dynamical mechanisms and selection effects.
Classical physics removed from the fundamental law category also 
the initial conditions for the electromagnetic field and 
all other forms of matter and energy (responsible for almost all the complexity we observe), 
leaving the fundamental laws quite simple.
A prospective theory of everything (TOE) incorporating a landscape of solutions populated by inflation reclassifies important aspects of the remaining ``laws'' as initial conditions. 
Indeed, those laws can differ from one post-inflationary region to another, and since 
inflation generically makes each such region enormous, its inhabitants might be fooled 
into misinterpreting regularities holding within their particular region as Universal (that is, multiversal) laws.
Finally, if the Level IV multiverse of all mathematical structures \cite{toe} exists, 
then even the ``theory of everything'' equations that physicists are seeking are merely local bylaws
in Rees' terminology \cite{ReesHabitat}, that vary across a wider ensemble.
Despite such retreats from {\it ab initio\/} explanations of certain phenomena, physics has progressed enormously in explanatory power.
}
\end{figure}

\section{Introduction}

Although the standard models of particle physics and cosmology 
have proven spectacularly successful, they
together require 31 free parameters
(Table~\FundParTable).  Why we observe them to have these particular values is an outstanding question in physics.

\subsection{Dimensionless numbers in physics}

This parameter problem can be viewed as the logical continuation of 
the age-old reductionist quest for simplicity.
Realization that the material world of chemistry and biology is 
built up from a modest number of elements entailed a 
dramatic simplification. But the observation of nearly one hundred chemical elements, more isotopes, and countless excited states eroded this simplicity. 

The modern $SU(3)\times SU(2)\times U(1)$ standard model of particle physics provides a much more sophisticated reduction.   Key properties (spin, electroweak and color charges) of  quarks, leptons and gauge bosons appear as labels describing  
representations of space-time and internal symmetry groups. 
The remaining complexity is encoded in 26 dimensionless numbers in the Lagrangian 
(Table~\FundParTable)\footnote{Here $\mu^2$ and $\lambda$ 
are defined so that the Higgs potential is $V(\Phi)=\mu^2|\Phi|^2+\lambda|\Phi|^4$.
}. 
All current cosmological observations can be fit with 5 additional parameters, though it is widely anticipated that
up to 6 more may be needed to accommodate more refined observations (Table~\FundParTable).

Table~{\DerivedParTable} expresses some common quantities in terms of these 31 fundamental ones\footnote{The last six
entries are mere order-of-magnitude estimates \cite{Weisskopf75,CarrRees79}.
In the renormalization group approximation for $\alpha$ in Table~{\DerivedParTable}, only fermions with mass below $m_Z$ should be included. 
}, with $\sim$ denoting cruder approximations than $\approx$.
Many other quantities commonly referred to as parameters or constants (see Table~{\DerivedQuantTable} for a sample) are not stable characterizations of properties 
of the physical world, since they vary markedly with time \cite{Scott05}.  
For instance, the baryon density parameter $\Omega_b$, the baryon density $\rho_b$, the Hubble parameter $h$ and
the cosmic microwave background temperature $T$ all decrease toward zero as the Universe 
expands and are, de facto, alternative time variables.

\begin{table*}
\noindent 
{\footnotesize
Table~\FundParTable: 
Input physical parameters.\footnote{Throughout this paper, we use extended Planck units $c=\hbar=G=k_b=|q_e|=1$.
For reference, this common convention gives the following units:
\begin{center}
\begin{tabular}{|l|l|l|}
\hline
Constant		&Definition		&Value\\
\hline
Planck length		&$(\hbar G/c^3)^{1/2}$  				&$1.61605\tento{-35}\Meter$\\
Planck time		&$(\hbar G/c^5)^{1/2}$  				&$5.39056\tento{-44}\Second$\\
Planck mass		&$(\hbar c/G)^{1/2}$					&$2.17671\tento{-8}\kg$\\
Planck temperature	&$(\hbar c^5/G)^{1/2}/k$				&$1.41696\tento{32}$K\\
Planck energy		&$(\hbar c^5/G)^{1/2}$  				&$1.22105\tento{19}$GeV\\ 
Planck density  	&$c^5/\hbar G^2$					&$5.15749\tento{96}\kg/\Meter^3$\\
Unit charge		&$|q_e|$						&$1.60218\times 10^{-19}\,$C\\
Unit voltage		&$(\hbar c^5/G)^{1/2}/|q_e|$				&$1.22105\tento{28}\,$V\\    
\hline
\end{tabular}
\end{center}     
} 
Those for particle physics (the first 26) and $\rhol$ appear explicitly in the the Lagrangian, 
whereas the cosmological ones (the last 11) are inserted as initial conditions. The last six are currently optional, but may
become required to fit improved measurements. The values are computed from the compilations in \cite{PDG,Mohapatra04,inflation}.
\begin{center}
{
\begin{tabular}{|l|ll|}
\hline
Parameter		&Meaning				 &Measured value\\
\hline
$g		$	&Weak coupling constant at $m_Z$	 &$0.6520\pm 0.0001$\\  
$\theta_W	$	&Weinberg angle				 &$0.48290\pm 0.00005$\\  
$g_s		$	&Strong coupling constant at $m_Z$	 &$1.221\pm 0.022$\\ 
\cline{2-3}
$\mu^2		$	&Quadratic Higgs coefficient		 &$\sim -10^{-33}$\\ 
$\lambda	$	&Quartic Higgs coefficient		 &$\sim 1$?\\ 
\cline{2-3}
$G_e            $	&Electron Yukawa coupling		 &$2.94\times 10^{-6}$\\
$G_\mu          $	&Muon Yukawa coupling			 &$0.000607$\\
$G_\tau         $	&Tauon Yukawa coupling			 &$0.0102156233$\\
\cline{2-3}
$G_u            $	&Up quark Yukawa coupling		 &$0.000016\pm 0.000007$\\
$G_d            $	&Down quark Yukawa coupling		 &$0.00003\pm 0.00002$\\
$G_c            $	&Charm quark Yukawa coupling		 &$0.0072\pm 0.0006$\\ 
$G_s            $	&Strange quark Yukawa coupling		 &$0.0006\pm 0.0002$\\
$G_t            $	&Top quark Yukawa coupling		 &$1.002\pm 0.029$\\ 
$G_b            $	&Bottom quark Yukawa coupling		 &$0.026\pm 0.003$\\
\cline{2-3}
$\sin\theta_{12}$	&Quark CKM matrix angle 		 &$0.2243\pm 0.0016$\\
$\sin\theta_{23}$	&Quark CKM matrix angle 		 &$0.0413\pm 0.0015$\\
$\sin\theta_{13}$	&Quark CKM matrix angle 		 &$0.0037\pm 0.0005$\\
$\delta_{13}	$	&Quark CKM matrix phase 		 &$1.05\pm 0.24$\\
\cline{2-3}
$\thetaqcd	$	&CP-violating QCD vacuum phase 		 &$<10^{-9}$\\
\cline{2-3}
$G_{\nu_e}	$      &Electron neutrino Yukawa coupling  	 &$<1.7\times 10^{-11}$\\	
$G_{\nu_\mu}	$      &Muon neutrino Yukawa coupling		 &$<1.1\times 10^{-6}$\\	
$G_{\nu_\tau}	$      &Tau neutrino Yukawa coupling		 &$<0.10$\\			
\cline{2-3}
$\sin\theta_{12}'$	&Neutrino MNS matrix angle 		 &$0.55\pm 0.06$\\
$\sin 2\theta_{23}'$	&Neutrino MNS matrix angle 		 &$\ge 0.94$\\
$\sin\theta_{13}'$	&Neutrino MNS matrix angle 		 &$\le 0.22$\\
$\delta_{13}'	$	&Neutrino MNS matrix phase 		 &$?$\\
\hline   
$\rhol		$	&Dark energy density						&$(1.25\pm 0.25)\tento{-123}$\\ 
$\xib		$	&Baryon mass per photon $\rhob/\ng$				&$(0.50\pm 0.03)\times 10^{-28}$\\
$\xic		$	&Cold dark matter mass per photon $\rhoc/\ng$			&$(2.5\pm 0.2)\times 10^{-28}$\\
$\xin		$	&Neutrino mass per photon $\rhon/\ng={3\over 11}\sum m_{\nu_i}$	&$<0.9\times 10^{-28}$\\ 
$Q       	$	&Scalar fluctuation amplitude $\delta_H$ on horizon		&$(2.0\pm 0.2)\times 10^{-5}$\\ 
\hline
$\ns	        $	&Scalar spectral index					        &$0.98\pm 0.02$\\ 
$\al 		$       &Running of spectral index $d\ns/d\ln k$		        &$|\alpha|\simlt 0.01$\\ 
$r           	$	&Tensor-to-scalar ratio	$(Q_t/Q)^2$      	        	&$\simlt 0.36$\\ 
$\nt           	$ 	&Tensor spectral index						&Unconstrained\\
$w		$	&Dark energy equation of state					&$-1\pm 0.1$\\ 
$\curvature     $	&Dimensionless spatial curvature $k/a^2 T^2$ \cite{Q}		&$|\curvature|\simlt 10^{-60}$\\ 
\hline
\end{tabular}
} 
\end{center}     
} 
\end{table*}

\begin{table*}
\noindent 
{\footnotesize
Table~\DerivedParTable: Derived physical parameters, in extended Planck units $c=\hbar=G=k_b=|q_e|=1$.
\begin{center}
\begin{tabular}{|l|lll|}
\hline
Parameter		&Meaning				&Definition		&Measured value\\
\hline   
$e            	$	&Electromagnetic coupling constant at $m_Z$ 	&$g\sin\theta_W$						&$\approx 0.313429\pm 0.000022$\\  
$\alpha(m_z)	$	&Electromagnetic interaction strength at $m_Z$	&$e^2/4\pi=g^2\sin^2\theta_W/4\pi$				&$\approx 1/(127.918\pm 0.018)$\\ 
$\alpha_w	$	&Weak interaction strength at $m_Z$		&$g^2/4\pi$							&$\approx 0.03383\pm 0.00001$\\ 	
$\alpha_s	$	&Strong interaction strength at $m_Z$		&$g_s^2/4\pi$							&$\approx 0.1186\pm 0.0042$\\ 
$\alpha_g	$	&Gravitational coupling constant		&$G\mp^2/\hbar c=\mp^2$						&$\approx 5.9046\tento{-39}$\\
$\alpha$	&	Electromagnetic interaction strength at $0$	&$\sim\left[\alpha(m_z)^{-1} + {2\over 9\pi}\ln{m_z^{20}\over m_u^4 m_c^4 m_d m_s m_b  m_e^3 m_\mu^3 m_\tau^3}\right]^{-1}$				&$1/137.03599911(46)$\\
$m_W            $	&$W^{\pm}$ mass				&$v g/2$							&$(80.425\pm 0.038)\GeV$\\
$m_Z            $	&$Z$ mass				&$v g/2\cos\theta_W$						&$(91.1876\pm 0.0021)\GeV$\\
$G_F            $	&Fermi constant				&$1/\sqrt{2}v^2$						&$\approx 1.17\times 10^{-5}\GeV^{-2}$\\
\cline{2-4}
$m_H		$	&Higgs mass				&$\sqrt{-\mu^2/2}$						&100-250 GeV?\\ 
$v		$	&Higgs vacuum expectation value		&$\sqrt{-\mu^2/\lambda}$					&$(246.7\pm 0.2)\GeV$\\  
\cline{2-4}
$m_e            $	&Electron mass				&$v G_e/\sqrt{2}$						&$(510998.92\pm 0.04)\eV$\\
$m_\mu          $	&Muon mass				&$v G_\mu/\sqrt{2}$						&$(105658369\pm 9)\eV$\\
$m_\tau         $	&Tauon mass				&$v G_\tau/\sqrt{2}$						&$(1776.99\pm 0.29)\MeV$\\
$m_u            $	&Up quark mass				&$v G_u/\sqrt{2}$						 &$(1.5-4)\MeV$\\
$m_d            $	&Down quark mass			&$v G_g/\sqrt{2}$						 &$(4-8)\MeV$\\
$m_c            $	&Charm quark mass			&$v G_c/\sqrt{2}$						 &$(1.15-1.35)\GeV$\\
$m_s            $	&Strange quark mass			&$v G_s/\sqrt{2}$						 &$(80-130)\MeV$\\
$m_t            $	&Top quark mass				&$v G_t/\sqrt{2}$						 &$(174.3\pm 5.1)\GeV$\\
$m_b            $	&Bottom quark mass			&$v G_b/\sqrt{2}$						 &$(4.1-4.9)\GeV$\\
$m_{\nu_e}	$      &Electron neutrino mass  		&$v G_{\nu_e}/\sqrt{2}$						&$<3\eV$\\
$m_{\nu_\mu}	$      &Muon neutrino mass			&$v G_{\nu_\mu}/\sqrt{2}$					&$<0.19\MeV$\\
$m_{\nu_\tau}	$      &Tau neutrino mass			&$v G_{\nu_\tau}/\sqrt{2}$					&$<18.2\GeV$\\
\cline{2-4}
$\mp            $	&Proton mass				&$2m_u+m_d+$QCD$+$QED						&$(938.27203\pm 0.00008)\MeV$\\
$\mn            $	&Neutron mass				&$2m_d+m_u+$QCD$+$QED						&$(939.56536\pm 0.00008)\MeV$\\
\cline{2-4}
$\beta		$	&Electron/proton mass ratio		&$m_e/m_p$							&$1/1836.15$\\ 
$\beta_n	$	&Neutron/proton relative mass difference&$m_n/m_p-1$							&$1/725.53$\\ 
\hline   
$\Ry        	$	&Hydrogen binding energy (Rydberg)	&$\Ry=m_e c^2\alpha^2/2=\alpha^2\beta\mp/2$			&$\approx 13.6057\eV$\\
$\aB    	$	&Bohr radius				&$\aB=\hbar/c m_e\alpha = (\alpha\beta\mp)^{-1}$	&$\approx 5.29177\tento{-11}\Meter$\\
$\st       	$	&Thomson cross section			&${8\pi\over 3} \left({\hbar\alpha\over m_e c}\right)^2 = {8\pi\over 3}(\alpha/\mp\beta)^2$&$\approx 6.65246\tento{-29}\,\Meter^2$\\
$k_c       	$	&Coulomb's constant			&$1/4\pi\epsilon_0 = \hbar c\alpha/q_e^2=\alpha 		$&$1/137.03599911(46)$\\
\hline
$\eta		$	&Baryon/photon ratio			&$\nb/\ng=\xib/m_p$						&$(6.3\pm 0.3)\tento{-10}$\\ 
$\xi		$	&Matter per photon 			&$\xib+\xic+\xin=\rhom/\ng=\mp\eta(1+\Rc+\Rn)$			&$(3.3\pm 0.3)\tento{-28}\approx 4$eV\\
$\Rc		$	&CDM/baryon density ratio		&$\rhoc/\rhob=\xic/\xib=\ocdm/\ob$				&$\approx 6$\\
$\Rn		$	&Neutrino/baryon density ratio		&$\rhon/\rhob={\fn\om\over\ob}={3\over 11}{\Mnu\over\eta\mp}$	&$\simlt 1$\\
$\Mnu    	$	&Sum of neutrino masses 		&$\Mnu=3\rhon/\nnu=(11/3)\mp\eta\Rn$			&\\
$\fnu		$	&Neutrino density fraction		&$\fn=\rhon/\rhom=\left[1+{11(\xib+\xid)\over 3\Mnu}\right]^{-1}$&$<0.1$\\ 
$\Teq	        $	&Matter-radiation equality temperature&${30\zeta(3)\over\pi^4}{\left[1+{21\over 8}\left({4\over 11}\right)^{4/3}\right]^{-1}}\xi\approx 0.220189\xi$ 	&$\approx 9.4\tento{3}$K\\
$\rhomeq	$	&Matter density at equality&		$\rhom(\Teq)=\frac{765314352000\zeta(3)^4}{{\left(242 + 21\times 22^{2/3}\right)}^3{\pi^{14}}}\xi^4 \approx 0.00260042\xi^4$ 		&$\approx 1.1\tento{-16}\kg/\Meter^3$\\  
$\Avac	        $	&Dark energy domination epoch	&$\xeq^{1/3}=(\rhomeq/\rhol)^{1/3}\approx 0.137514\xi^{4/3}\rhol^{-1/3}$	&$3215\pm 639$\\
\hline   
$m_{\rm galaxy}$       	&Mass of galaxy \cite{CarrRees79}	      		&$\sim \alpha^5\beta^{-1/2}\mp^{-3}$			&$\sim 10^{41}\kg$\\ 
$\mstar$		&Mass of star \cite{CarrRees79}				&$\sim\mp^{-2}$						&$\sim 10^{30}\kg$\\
$m_{\rm planet}$       	&Mass of habitable planet \cite{CarrRees79}	      	&$\sim 10^{-4}\alpha^{3/2}\mp^{-2}$			&$\sim 10^{24}\kg$\\ 
$m_{\rm asteroid}$      &Maximum mass of asteroid \cite{CarrRees79}	      	&$\sim\alpha^{3/2}\beta^{3/2}\mp^{-1/2}$		&$\sim 10^{22}\kg$\\ 
$m_{\rm person}$       	&Maximum mass of person \cite{CarrRees79}	      	&$\sim 10^2\alpha^{3/4}\mp^{-1/2}$			&$\sim 10^2\kg$\\ 
$\xiwimp$      		&WIMP dark matter density per photon			&$\sim 10^5 v^2/g^2=-10^5\mu^2/\lambda g^2$		&$\sim 3\times 10^{-28}?$\\
\hline   
\end{tabular}
\end{center}     
} 
\end{table*}

\begin{table*}
\noindent 
{\footnotesize
Table~\DerivedQuantTable: Derived physical variables, in extended Planck units $c=\hbar=G=k_b=|q_e|=1$.
\begin{center}
\begin{tabular}{|l|lll|}
\hline
Parameter		&Meaning				&Definition		&Measured value\\
\hline
$T		$	&CMB temperature		&(Acts as a time variable)						&$(2.726\pm 0.005)$K (today)\\ 
$\ng       	$	&Photon number density		&${2\zeta(3)\over\pi^2} T^3,\quad \zeta(3)\approx 1.20206$		&$0.243588 T^3$\\
$\nnu       	$	&Neutrino number density	&${9\over 11}\ng = {18\zeta(3)\over 11\pi^2} T^3$			&$0.199299 T^3$\\
$\rhog       	$	&Photon density			&${\pi^2\over 15} T^4$							&$0.657974 T^4$\\
$\rhob		$	&Baryon density			&$\xib\ng=m_p\eta\ng={2\zeta(3)\over\pi^2}\xib T^3$			&$0.243588\xib T^3$\\
$\rhoc		$	&CDM density			&$\xic\ng=\Rc\rhob={2\zeta(3)\over\pi^2}\xic T^3$			&$0.243588\xic T^3$\\
$\rhon		$	&Neutrino density (massive)	&$\xin\ng=\Rn\rhob={2\zeta(3)\over\pi^2}\xin T^3={\nnu\Mnu\over 3}={3\over 11}\ng\Mnu$	&$0.243588\xin T^3$\\
$\rhong       	$	&Neutrino density (massless)	&${21\over 8}\left({4\over 11}\right)^{4/3}\rhog\approx 0.681322\rhog$	&$0.448292 T^4$\\
$\rhom		$	&Total matter density		&$\rhob+\rhoc+\rhon=\ng\xi={2\zeta(3)\over\pi^2}(\xib+\xic+\xin)T^3=\olam\x$&$0.243588\xi T^3$\\
$\rhok		$	&Curvature density		&$-{3k\over 8\pi a^2}=-{3\over 8\pi}\curvature T^2$			&$-0.119366\curvature T^2$\\
$A	        $	&Expansion factor since equality&$a/\aeq=\Avac x^{1/3}\approx 0.137514 \xi^{4/3}\rhom^{-1/3}$		&$\approx 3.5\tento{3}$ today\\
$x	        $	&Dark energy/matter ratio	&$\rhol/\rhom = (A/\Avac)^3 = {\pi^2\over 2\zeta(3)}{\rhol\over\xi T^3}$&$\approx 7/3$ today\\
\hline
$\Hstar		$	&Hubble reference rate (mere unit)&$H/h = 100\km\>\Second^{-1}\Mpc^{-1}$					&$(9.7779\>\Gyr)^{-1}$\\
$\rhoh	        $	&Hubble reference density (mere unit)&$3\Hstar^2/8\pi G = 3(100\km\>\Second^{-1}\Mpc^{-1})^2/8\pi G$	&$1.87882\tento{-26}\kg/\Meter^3$\\
$\ob   		$       &Baryon density parameter	&$\Ob h^2 = \rhob/\rhoh = \xib\ng/\rhoh = (2\zeta(3)/\pi^2)\xib T^3$	&$0.023\pm 0.001$ today\\ 
$\ocdm   	$       &Cold dark matter density parameter&$\Oc h^2 =\rhoc/\rhoh = \xic\ng/\rhoh = (2\zeta(3)/\pi^2\rhoh)\xic T^3$	&$0.12\pm 0.01$ today\\ 
$\on   		$       &Neutrino density parameter	&$\On h^2 = \rhon/\rhoh = \xin\ng/\rhoh = (2\zeta(3)/\pi^2\rhoh)\xin T^3$	&$<0.01$ today\\ 
$\od   		$       &Dark matter density parameter	&$\Od h^2 = \ocdm+\on = (2\zeta(3)/\pi^2\rhoh)(\xib+\xic) T^3$			&$\approx 0.023\pm 0.001$ today\\ 
$\om   		$       &Matter density parameter	&$\Om h^2 = \ob+\ocdm+\on =(2\zeta(3)/\pi^2\rhoh)(\xib+\xic+\xin) T^3$	&$0.14\pm 0.01$ today\\
$\olam  	$       &Dark energy density parameter	&$\Ol h^2 = \rhol/\rhoh$							&$0.34\pm 0.09$\\ 
$\og  		$       &Photon density	parameter	&$\rhog/\rhoh= (\pi^2/15\rhoh) T^4 \approx 1.80618\times 10^{122}T^4$		&$0.0000247\pm 0.0000004$ today\\
$H	        $	&Hubble parameter		&$\left[{8\pi G\over 3}(\rhol+\rhom+\rhog+\rhok)\right]^{1/2}$		&$\approx (10\>\Gyr)^{-1}$ today\\
$h	        $	&Dimensionless Hubble parameter	&$H/H_*$								&$\approx 0.7$ today\\
$t    		$	&Age of Universe		&$\int_0^a H(a')^{-1} d\ln a'$						&$\approx 14\,\Gyr$ today\\
$\trd		$	&Age during radiation era	&${3\over 4\pi}\sqrt{3\over 2}\left[1+{21\over 8}\left({4\over 11}\right)^{4/3}\right]^{-1/2}T^{-2}$	&$\approx 0.225492/T^2$\\ 
$\tmd		$	&Age during matter era		&$\approx 1/[12\zeta(3)]^{1/2}\xi^{1/2} T^{3/2}$			&$\approx 0.2633\xi^{-1/2} T^{-3/2}$\\
$\tvd		$	&Age during vacuum era		&$\approx {3\over 8\pi\rhol}\ln{\xi^{1/3} T\over\rhol^{1/3}}$			&\\
$\Ol		$	&Vacuum density ratio		&$\olam/h^2=\rhol/h^2\rhoh$ 							&$\approx 0.7$ today\\
$\Om		$	&Matter density ratio		&$\om/h^2$ (analogously, $\Ob\equiv\ob/h^2$, $\Od\equiv\od/h^2$, \etc). 	&$\approx 0.3$ today\\
$\Otot       	$	&Spatial curvature parameter	&$1+\curvature (T/H)^2$								&$1.01\pm 0.02$ today\\ 
$\As		$	&Scalar power normalization 	&$[(3/2\pi)(T/1\mK) Q]^2/800$ 							&$\approx 0.8$ today\\
\hline   
\end{tabular}
\end{center}     
} 
\end{table*}

Our particular choice of parameters in Table~{\FundParTable} is a compromise balancing simplicity of expressing the fundamental laws (i.e., the Lagrangian of the standard model and the equations for cosmological evolution) and ease of measurement.   All parameters except $\mu^2$, $\rhol$, $\xib$, $\xic$ and $\xin$ are intrinsically dimensionless, and we make
these final five dimensionless by using Planck units (for alternatives, see \cite{WilczekUnits1,WilczekUnits2}). 
Throughout this paper, we use ``extended'' Planck units
defined by $c=G=\hbar=|q_e|=k_B=1$.  We use $\hbar=1$ rather than $h=1$ to minimize the number of $(2\pi)$-factors elsewhere.

\subsection{The origin of the dimensionless numbers}

So why do we observe these 31 parameters to have the particular values listed in Table~{\FundParTable}?
Interest in that question has grown with the gradual realization that some of these parameters 
appear fine-tuned for life, in the sense that 
small relative changes to their values would result in dramatic qualitative changes that could preclude intelligent life, and hence the very possibility of reflective observation.
As discussed extensively elsewhere \cite{Carter74,BarrowTipler,ReesHabitat,BostromBook,multiverse,conditionalization,endofanthropicprinciple,Davies04,Hogan04,Stoeger04,Aguirre05,LivioRees05,Weinstein05,Weinberg05},
there are four common responses to this realization:
\vskip1cm
\begin{enumerate}
\item {\bf Fluke:} Any apparent fine-tuning is a fluke and is best ignored.
\item {\bf Multiverse:} These parameters vary across an ensemble of physically realized and (for all practical purposes) parallel universes, 
and we find ourselves in one where life is possible. 
\item {\bf Design:} Our universe is somehow created or simulated with parameters chosen to allow life.
\item {\bf Fecundity:} There is no fine-tuning, because intelligent life of some form will emerge under extremely varied circumstances.
\end{enumerate}
Options 1, 2, and 4 
tend to be preferred by physicists, with recent
developments in inflation and high-energy theory giving new popularity to option 2.

Like relativity theory and quantum mechanics, the theory of inflation has not only
solved old problems, but also widened our intellectual horizons, arguably deepening our understanding of the nature of physical reality.
First of all, inflation is generically eternal
\cite{LindeBook,Vilenkin83,Starobinsky84,Starobinsky86,Goncharov86,SalopekBond91,LindeLindeMezhlumian94},
so that even though inflation has ended in the part of space that we inhabit, it still continues elsewhere and
will ultimately produce an infinite number of post-inflationary volumes as large as ours, forming a cosmic fractal of sorts.
Second, these regions may have different physical properties. 
This can occur in fairly conventional contexts involving symmetry breaking, without invoking more exotic aspects of eternal inflation.  
(Which, for example, might also allow different values of the axion dark matter density parameter $\xic$
\cite{Linde88,Wilczek04,Linde04}.)
More dramatically, a common feature of much string theory related model building is that 
there is a ``landscape'' of solutions, corresponding to spacetime configurations involving 
different values of both seemingly continuous parameters (Table~\FundParTable)
and discrete parameters (specifying the spacetime dimensionality, the gauge group/particle content, \etc), 
some or all of which may vary across the landscape \cite{Bousso00,Feng00,KKLT03,Susskind03,AshikDouglas04}.

If correct, eternal inflation might transform that potentiality into reality, actually creating regions of space realizing each of
these possibilities. Generically each region where inflation has ended is infinite in size, therefore potentially fooling 
its inhabitants into mistaking initial conditions for fundamental laws. 
Inflation may thus indicate the same sort of shift in the borderline between fundamental and effective laws of physics
(at the expense of the former) previously seen in theoretical physics, as illustrated in \fig{BoundaryFig}.

There is quite a lot that physicists might once have hoped to derive from fundamental principles,
for which that hope now seems naive and misguided \cite{WilczekPT04}.
Yet it is important to bear in mind that these 
philosophical retreats have gone hand in hand with massive progress in predictive power. 
While Kepler and Newton discredited {\it ab initio} attempts to explain planetary orbits and shapes with circles and spheres being ``perfect shapes'',
Kepler enabled precise predictions of planetary positions, and 
Newton provided a dynamical explanation of the approximate sphericity of planets and stars.
While classical physics removed all initial conditions from its predictive purview, its explanatory power inspired awe.
While quantum mechanics dashed hopes of predicting when a radioactive atom would decay,
it provided the foundations of chemistry, and it predicts a wealth of surprising new phenomena, as we continue to discover.

\subsection{Testing fundamental theories observationally}

Let us group the 31 parameters of {Table~\FundParTable} into a 31-dimensional vector $\p$. 
In a fundamental theory where inflation populates a landscape of possibilities,
some or all of these parameters will vary from place to place as described
by a 31-dimensional probability distribution $f(\p)$.  Testing this theory observationally
corresponds to confronting that theoretically predicted distribution with the values we observe.
Selection effects make this challenging \cite{Carter74,BostromBook}: if any of the parameters that can vary affect
the formation of (say) protons, galaxies or observers, then the parameter probability distribution differs depending 
on whether it is computed at a random point, a random proton, a random galaxy or a random observer \cite{BostromBook,conditionalization}.
A standard application of 
conditional probabilities predicts the observed distribution 
\beq{BayesEq}
f(\p)\propto\fprior(\p)\fselec(\p),
\eeq
where $\fprior(\p)$ is the theoretically predicted distribution at a random point at the end of inflation
and $\fselec(\p)$ is the probability of our observation being made at that point.
This second factor $\fselec(\p)$, incorporating the selection effect, is simply proportional to 
the expected number density of reference objects formed (say, protons, galaxies or observers).

Including selection effects when comparing theory against observation is
no more optional than the correct use of logic.   
Ignoring the second term in \eq{BayesEq} can even reverse the verdict as to whether 
a theory is ruled out or consistent with observation.
On the other hand, anthropic arguments that ignore the first term in \eq{BayesEq}
are likewise spurious;  it is crucial to know which of the parameters can vary, how these variations are
correlated, and whether the typical variations are larger or smaller than constraints arising from
the selection effects in the second term.

\subsection{A case study: cosmology and dark matter}

Examples where we can compute {\it both} terms in \eq{BayesEq} are hard to come by.
Predictions of fundamental theory for the first term, insofar as they are plausibly formulated at present, tend to take the form of 
functional constraints among the parameters.  Familiar examples are the constraints among couplings arising from gauge 
symmetry unification and the constraint $\theta_{\rm QCD} \approx 0$ arising from Peccei-Quinn symmetry.  
Attempts to predict the distribution of inflation-related cosmological parameters are marred at present by 
regularization issues related to comparing infinite volumes 
\cite{LindeMezhlumian93,LindeLindeMezhlumian94,Garcia94,Garcia95,LindeLindeMezhlumian,Vilenkin95,WinitzkiVilenkin96,LindeMezhlumian96,LindeLindeMezhlumian96,VilenkinWinitzki97,Vilenkin98,Vanchurin00,Garriga01,Garriga05}.
Additional difficulties arise from our limited understanding as to what to count as an observer, when we consider variation in parameters that affect the evolution of life, such as 
$(m_p,\alpha,\beta)$, which approximately determine all properties of chemistry.

In this paper, we will focus on a rare example where 
where there are no problems of principle in computing {\it both\/} terms:
that of cosmology and dark matter,
involving variation in the parameters $(\xic,\rhol,Q)$ from Table~\FundParTable, \ie, the
dark matter density parameter, the dark energy density and the seed fluctuation amplitude.
Since none of these three parameters affect the evolution of life at the level of biochemistry,
the only selection effects we need to consider
are astrophysical ones related to the formation of dark matter halos, galaxies and stable solar systems.
Moreover, as discussed in the next section, we have specific well-motivated 
prior distributions for $\rhol$ and (for the case of axion dark matter) $\xic$.
Making detailed dark matter predictions is interesting and timely given the major efforts underway to detect 
dark matter both directly \cite{Irastorza05}
and indirectly \cite{Pacheco05} and the prospects of discovering supersymmetry and 
a WIMP dark matter candidate in Large Hadron Collider operations from 2008.

For simplicity, we do not include any of the currently optional cosmological parameters, 
\ie, we take $\ns=\Otot=1$, $\al=r=\nt=0$, $w=-1$.
The remaining two non-optional cosmological parameters in Table~{\FundParTable} are
the density parameters $\xin$ for neutrinos and $\xib$ for baryons.
It would be fairly straightforward to generalize our treatment below to include
$\xin$ along the lines of \cite{anthroneutrino,anthrolambdanu}, since it too affects $\fselec$ only through astrophysics
and not through subtleties related to biochemistry. Here, for simplicity, we will ignore it; in any case, it
has been observed to be rather unimportant cosmologically ($\xin\ll\xic$).
When computing cosmological fluctuation growth, we will also make the simplifying approximation 
that $\xib\ll\xic$ (so that $\xi\sim\xic$), although we will include $\xib$ as a free parameter 
when discussing galaxy formation and solar system constraints. 
(For very large $\xib$, structure formation can change qualitatively; see~\cite{Aguirre99}.)
We will see below that $\xic/\xib\gg 1$ is not only observationally indicated (Table~{\FundParTable} gives 
$\xic/\xib\sim 6$), but also emerges as the theoretically most interesting regime if $\xib$ is considered fixed.

The rest of this paper is organized as follows.
In \Sec{PriorSec}, we discuss theoretical predictions for the first term of \eq{BayesEq}, the prior distribution $\fprior$.
In \Sec{SelectionSec}, we discuss the second term $\fselec$, 
computing the selection effects corresponding to halo formation, galaxy formation and 
solar system stability.
We combine these results and make predictions for dark-matter-related parameters in \Sec{ResultsSec}, 
summarizing our conclusions in \Sec{ConclusionsSec}.
A number of technical details are relegated to Appendix A.

\section{Priors}
\label{PriorSec}

In this section, we will discuss the first term in \eq{BayesEq}, specifically how the function $\fprior(\p)$
depends on the parameters $\xic$, $\rhol$ and $Q$. In the case of $\xic$, we will consider two dark matter candidates, axions and WIMPs.
 
\subsection{Axions}
\label{AxionSec}

The axion dark matter model offers an elegant example where the 
prior probability distribution of a parameter (in this case $\xic$) can be computed analytically.

The strong CP problem is the fact that the dimensionless parameter $\thetaqcd$ 
in Table~{\FundParTable}, which parameterizes a potential CP-violating term in quantum chromodynamics (QCD), the theory of the strong interaction, is so small.   Within the standard model, $\thetaqcd$ is a periodic variable whose possible values run from 0 to $2\pi$, so its natural scale is of order unity.   Selection effects are of little help here, since values of $\thetaqcd$ far larger than the observed bound $|\thetaqcd|\simlt 10^{-9}$ would seem to have no serious impact on life.

Peccei and Quinn \cite{PecceiQuinn} introduced microphysical models that address the strong CP problem.  Their models extend the standard model so as to support an appropriate (anomalous, spontaneously broken) symmetry.   The symmetry is called Peccei-Quinn (PQ) symmetry, and the energy scale at which it breaks is called the Peccei-Quinn scale.  Weinberg \cite{Weinberg78} and Wilczek \cite{Wilczek78} independently realized that Peccei-Quinn symmetry implies the existence of a field whose quanta are extremely light, extremely feebly interacting particles, known as axions.  
Later it was shown that axions provide an interesting dark matter candidate
\cite{PreskillWiseWilczek83,AbbottSikivie83,DineFischler83}.

Major aspects of axion physics can be understood by reference to a truncated toy model where $\thetaqcd$ is the complex phase angle
of a complex scalar field $\Phi$ that develops a potential of the type
\beq{VaxionEq}
V(\Phi) = (|\Phi|^2-\fa^2)^2 + \Lambdaqcd^4(1-\fa^{-1}{\rm Re}\,\Phi),
\eeq
where $\Lambdaqcd\sim 200\MeV$ is ultimately determined by the parameters in Table~{\FundParTable}; roughly speaking, it is 
the energy scale where the strong coupling constant $\alphas(\Lambdaqcd)=1$.

At the Peccei-Quinn (PQ) symmetry breaking scale $\fa$, assumed to be much larger than $\Lambdaqcd$, 
this complex scalar field $\Phi$ feels a Mexican hat potential
and seeks to settle toward a minimum $\expec{\Phi}=\fa e^{i\thetaqcd}$.  In the context of cosmology, this will occur at temperatures not much below $\fa$.  Initially the angle $\thetaqcd=\theta_0$ is of negligible energetic significance, and so it is 
effectively a random field on superhorizon scales.  The angular part of this field is called the axion field $a=\fa\thetaqcd$.
As the cosmic expansion cools our universe to much lower temperatures approaching the QCD scale,
the approximate azimuthal symmetry of the Mexican hat is broken by the emergence of the second term,
a periodic potential $1-\fa^{-1}{\rm Re}\,\Phi=1-\cos\thetaqcd=2\sin^2\thetaqcd$ (induced by QCD instantons) 
whose minimum corresponds to no strong CP-violation, 
\ie, to $\thetaqcd=0$.  
The axion field oscillates around this minimum like an underdamped harmonic oscillator 
with a frequency $m_a$ corresponding to the second derivative of the potential at the minimum,
gradually settling towards this minimum as the oscillation amplitude is damped by Hubble friction.  That oscillating field can be interpreted as a Bose condensate of axions.  It obeys the equation of state of a low-pressure gas, which is to say it provides a form of cold dark matter.   

By today, $\thetaqcd$ is expected to have  settled to an angle within about $10^{-18}$ of its minimum \cite{DineFischler83},
comfortably below the observational limit $|\thetaqcd|\simlt 10^{-9}$, and thus dynamically solving the strong CP problem.  (The exact location of the minimum is model-dependent, and not quite at zero, but comfortably small in realistic models \cite{Pospelov}.)

The axion dark matter density per photon in the current epoch is estimated to be 
\cite{PreskillWiseWilczek83,AbbottSikivie83,DineFischler83}
\beq{axionDensityEq}
\xic=\xistar\sin^2{\theta_0\over 2},\quad \xistar\sim\fa^4.
\eeq
If axions constitute the cold dark matter and the Peccei-Quinn phase transition occurred well before the end of 
inflation, then the measurement $\xic\sim 3\times 10^{-28}$ thus implies that
\beq{faEq}
\fa\sim 10^{-7}\left(\sin{\theta_0\over 2}\right)^{-1/2} \sim 10^{12}\GeV\times\left(\sin{\theta_0\over 2}\right)^{-1/2},
\eeq
where $\theta_0$ is the the initial misalignment angle of the axion field in our particular Hubble volume.

Frequently it has been argued that 
this implies $\fa\sim 10^{12}\GeV$, ruling out GUT scale axions with $\fa\sim 10^{16}\GeV$.  Indeed, in a conventional cosmology the horizon size at the Peccei-Quinn transition corresponds to a small volume of the universe today, and the observed universe on cosmological scales would fully sample the random distribution $\theta_0$.   
However, the alternative possibility that $|\theta_0|\ll 1$ over our entire observable universe was pointed out already in 
\cite{PreskillWiseWilczek83}.   It can occur if an epoch of inflation intervened between Peccei-Quinn symmetry breaking and the present; in that case the observed universe arises from {\it within\/} a single horizon volume  
at the Peccei-Quinn scale, and thus plausibly lies within a correlation volume.   
Linde \cite{Linde88} argued that if there were an anthropic selection effect against
very dense galaxies, then models with $\fa\gg 10^{12}\GeV$ and $|\theta_0|\ll 1$ might indeed be perfectly reasonable.  Several additional aspects of this scenario were discussed in \cite{Wilczek04,Linde04}.
Much of the remainder of this paper arose as an attempt to better ground its astrophysical foundations, but most of our considerations are of much broader application.

We now compute the axion prior $\fprior(\xic)$.
Since the symmetry breaking is uncorrelated between causally disconnected regions, 
$\theta_0$ is for all practical purposes a random variable that varies with a uniform distribution
between widely separated Hubble volumes.
Without loss of generality, we can take the interval over which $\theta_0$ varies to 
be $0\le\theta_0\le\pi$.
This means that the probability of $\xi$ being lower than some given value $\xi_0\le\xi_*$ is
\beqa{xiProbEq}
P(\xi<\xi_0)&=&P(\xistar\sin^2{\theta_0\over 2}<\xi_0) = P\left(\theta_0<2\sin^{-1}{\xi_0^{1/2}\over\xistar^{1/2}}\right)\nonumber\\
	    &=&{2\over\pi}\sin^{-1}{\xi_0^{1/2}\over\xistar^{1/2}}.
\eeqa
Differentiating this expression with respect to $\xi_0$ gives
the prior probability distribution for the dark matter density $\xistar$:
\beq{axionDensityEq2}
f_\xi(\xi) = {1\over\pi\left(\xistar\xi\right)^{1/2}\left(1-{\xi\over\xistar}\right)}
\eeq
For the case at hand, we only care about the tail of the prior corresponding to unusually small $\theta_0$,
\ie, the case $\xi\ll\xistar$, for which the probability distribution reduces
to simply
\beq{axionDensityEq3}
f_\xi(\xi) \propto {1\over\sqrt{\xi}}.
\eeq
Although this may appear to favor low $\xi$, the probability per logarithmic interval $\propto\sqrt{\xi}$,
and it is obvious from \eq{xiProbEq} that the bulk of the probability lies near the very high value $\xi\sim\xistar$.

A striking and useful property of \eq{axionDensityEq3} is that it contains no free parameters whatsoever.
In other words, this axion dark matter model makes an unambiguous prediction for the prior distribution of 
one of our 31 parameters, $\xic$.
Since the axion density is negligible at the time of inflation, 
this prior is immune to the inflationary measure-related problems discussed in \cite{inflation},
and no inflation-related effects should correlate $\xic$ with other observable parameters.
Moreover, this conclusion applies for quite general axion scenarios, not merely for our toy model --- the 
only property of the potential used to derive the $\xic^{-1/2}$-scaling 
is its parabolic shape near any minimum.
Although many theoretical subtleties arise regarding the axion dark matter scenario
in the contexts of inflation, supersymmetry and string theory \cite{TurnerWilczek91,Linde91,BanksDine97,BanksDineGraesser03},
the $\xic^{-1/2}$ prior appears as a robust consequence of the hypothesis $\fa \gg 10^{12}\GeV$.

We conclude this section with a brief discussion of bounds from axion fluctuations.
Like any other massless field, the axion field $a=f_a\theta$ acquires fluctuations of order $H$   
during inflation, where $H\sim\Einfl^2/\mpl=\Einfl^2$ and $\Einfl$ is the inflationary energy scale, so
$\delta\theta_0\sim\Einfl^2/\fa$.
For our $|\theta_0|\ll  1$ case, \eq{axionDensityEq} gives $\theta_0\sim\xic^{1/2}/\fa^2$, so 
we obtain the axion density fluctuations amplitude
\beq{QaEq}
Q_a\equiv{\delta\xic\over\xic}\approx {2\delta\theta_0\over\theta_0}\sim {\Einfl^2\fa\over\xic^{1/2}}.
\eeq
Such axion isocurvature fluctuations (see, \eg, \cite{Burns97} for a review)
would contribute acoustic peaks in the cosmic microwave background (CMB) out of phase with the
those from standard adiabatic fluctuations, allowing an observational upper bound
$Q_a\simlt 0.3 Q\sim 10^{-5}$ to be placed \cite{Peiris03,Moodley04}.
Combining this with the observed $\xic$-value from Table~{\FundParTable} gives the bound
$\Einfl^2\fa\sim Q_a\xic^{1/2}\simlt 10^{-19}$, bounding the 
inflation scale.
The traditional value $\fa\sim 10^{12}\GeV\sim  10^{-7}$ gives
the familiar bound $\Einfl\simlt 10^{-6}\sim 10^{13}\GeV$ \cite{Burns97,Fox04}.
A higher $\fa$ gives a tighter limit on the inflation scale:
increasing $\fa$ to the Planck scale ($\fa\sim 1$) lowers the 
bound to $\Einfl\simlt 10^{-9}\sim 10^{10}\GeV$ --- the constraint grows stronger
because the denominator in $\delta\theta_0/\theta_0$ must be smaller to avoid an excessive axion density.

For comparison, inflationary gravitational waves have amplitude $rQ\sim H\sim\Einfl^2$, so they are unobservably
small unless $\Einfl\simgt 10^{16}\GeV$.
Although various loopholes to the axion fluctuation bounds have been proposed 
(see \eg, \cite{Burns97,Fox04,Yanagida97,Khlopov05}),
it is interesting to note that the simplest axion dark matter models therefore make the falsifiable prediction
that future CMB experiments will detect no gravitational wave signal \cite{Fox04}.

\subsection{WIMPs}
\label{WimpSec}

Another popular dark matter candidate is a stable weakly interacting massive particle (WIMP), 
thermally produced in the early universe and with its relic abundance set by a standard freezeout calculation.  See, \eg, 
\cite{Jungman95,Feng05} for reviews.
Stability could, for instance, be ensured by the WIMP being the lightest supersymmetric particle.

At relevant (not too high) temperatures the thermally averaged WIMP annihilation cross section takes the form \cite{Jungman95}
\beqa{wimpCrossSectionEq}
\expec{\sigma v}=\gamma{\alphaw^2\over\mwimp^2},
\eeqa
where $\gamma$ is a dimensionless constant of order unity.
In our notation, the WIMP number density is $\nwimp=\rhowimp/\mwimp=\ng\xiwimp/\mwimp$.
The WIMP freezout is determined by equating the WIMP annihilation rate 
$\Gamma\equiv\expec{\sigma v}\nwimp=\expec{\sigma v}\ng\xiwimp/\mwimp$
with the radiation-dominated Hubble expansion rate $H=(8\pi\rhog/3)^{1/2}$.
Solving this equation for $\xiwimp$ and substituting the expressions for $\alphaw$, $\ng$ and $\rhog$ from 
Tables~{\DerivedParTable} and~{\DerivedQuantTable} gives
\beq{xiwimpEq}
\xiwimp = \sqrt{2^9\pi^{11}\over 45} {\mwimp^3\over\zeta(3)\gamma g^4 T}\approx 1522 {\mwimp^3\over\gamma g^4 T}
\eeq
The WIMP freezeout temperature is typically found to be of order $T\sim\mwimp/20$ \cite{Jungman95}.
If we further assume that the WIMP mass is of order the electroweak scale
($\mwimp\sim v$) and that the annihilation cross section prefactor $\gamma\sim 1$, then \eq{xiwimpEq} gives
\beq{xiwimpEq2}
\xiwimp\sim 10^5 {\mwimp^2\over g^4}\sim 10^5 {v^2\over g^4}\sim 10^{-28}
\eeq
for the measured values of $v$ and $g$ from Table~{\FundParTable}.
This well-known fact that the predicted WIMP abundance agrees qualitatively with the measured 
dark matter density $\xic$ is a key reason for the popularity of WIMP dark matter.

In contrast to the above-mentioned axion scenario, we have no 
compelling prior for the WIMP dark matter density parameter $\xiwimp$.
Let us, however, briefly explore the interesting scenario advocated by, \eg, \cite{ArkaniHamed05},
where the theory prior determines all relevant standard model parameters except the Higgs vacuum expectation value 
$v$, which has a broad prior distribution.\footnote{If only two microphysical parameters from Table~{\FundParTable} vary
by many orders of magnitude across an ensemble and are anthropically selected, one might be tempted to guess that they are 
$\rhol\sim 10^{-123}$ and $\mu^2\sim -10^{-33}$, since they differ most dramatically from unity.
A broad prior for $-\mu^2$ would translate into a broad prior for $v$ that could potentially provide an anthropic solution
to the so-called ``hierarchy problem'' that $v\sim 10^{-17}\ll 1$.
}
(It should be said, however, that the connection $\mwimp\sim v$ is somewhat artificial in this context.)  It has recently been shown that $v$ is subject to quite strong microphysical selection effects that have nothing 
to do with dark matter, as nicely reviewed in \cite{Hogan00}.
As pointed out by \cite{Agrawal97,Agrawal98}, changing $v$ up or down from its observed value $v_0$ by a large factor would 
correspond to a dramatically less complex universe because the slight neutron-proton mass difference has a 
quark mass contribution
$(m_d-m_u)\propto v$ that slightly exceeds the 
extra Coulomb repulsion contribution to the proton mass:
\begin{enumerate}
\item For $v/v_0\simlt 0.5$, protons (uud) decay into neutrons (udd) giving a universe with no atoms.
\item For $v/v_0\simgt 5$, neutrons decay into protons even inside nuclei, giving a universe with no atoms except hydrogen.
\item For $v/v_0\simgt 10^3$, protons decay into $\Delta^{++}$ (uuu) particles, giving a universe with only Helium-like $\Delta^{++}$ atoms.
\end{enumerate}
Even smaller shifts would qualitatively alter the synthesis of heavy elements: 
For $v/v_0\simlt 0.8$, diprotons and dineutrons are bound, producing a universe devoid of, \eg, hydrogen.
For $v/v_0\simgt 2$, deuterium is unstable, drastically altering standard stellar nucleosynthesis.

Much stronger selection effects appear to result from carbon and oxygen production in stars.
Revisiting the issue first identified by Hoyle \cite{Hoyle54} with 
numerical nuclear physics and stellar nucleosynthesis calculations, \cite{Oberhummer00} quantified how
changing the strength of the nucleon-nucleon interaction altered the yield of carbon and oxygen in various types of stars. 
Combining their results with those of \cite{Jeltema99} that relate the relevant nuclear physics parameters to $v$ gives
the following striking results:
\begin{enumerate}
\item For $v/v_0\simlt 0.99$, orders of magnitude less carbon is produced.
\item For $v/v_0\simgt 1.01$, orders of magnitude less oxygen is produced.
\end{enumerate}
Combining this with \eq{xiwimpEq2}, we see that this could potentially translate into a percent level selection effect 
on $\xiwimp\propto v^2$.

In the above-mentioned scenario where $v$ has a broad prior whereas the other particle physics parameters 
(in particular $g$) do not \cite{ArkaniHamed05},
the fact that microphysical selection effects on $v$ are so sharp translates into a narrow probability distribution for
$\xiwimp$ via \eq{xiwimpEq2}.
As we will see, the astrophysical selection effects on the dark matter density parameter are 
much less stringent.
We should emphasize again, however, that this constraint relies on the assumption of a tight connection between 
$\xiwimp$ and $v$, which could be called into question.

\subsection{$\rhol$ and $Q$}

As discussed in detail in the literature 
(\eg, \cite{BarrowTipler,LindeLambda,Weinberg87,Efstathiou95,Vilenkin95,Martel98,GarrigaVilenkin03,inflation}),
there are plausible reasons to adopt a prior on $\rhol$ that is essentially constant and independent of other parameters
across the narrow range where $|\rhol|\simlt Q^3\xi^4\sim 10^{-123}$ where $\fselec$ is non-negligible (see \Sec{SelectionSec}).
The conventional wisdom is that since $\rhol$ is the difference between two much larger quantities, and $\rhol=0$ has no evident microphysical significance, 
no ultrasharp features appear in the probability distribution for $\rhol$
within $10^{-123}$ of zero.
That argument holds even if $\rhol$ varies discretely rather than continuously,
so long as it takes $\gg 10^{123}$ different values across the ensemble.   

In contrast, calculations of the prior distribution for $Q$ from inflation are fraught with considerable uncertainty
\cite{inflation,GarrigaVilenkin05,Linde05}.  We therefore avoid making assumptions about this function in our calculations.

\clearpage
\section{Selection effects}
\label{SelectionSec}

\begin{figure} 
\centerline{\epsfxsize=\figsize\epsffile{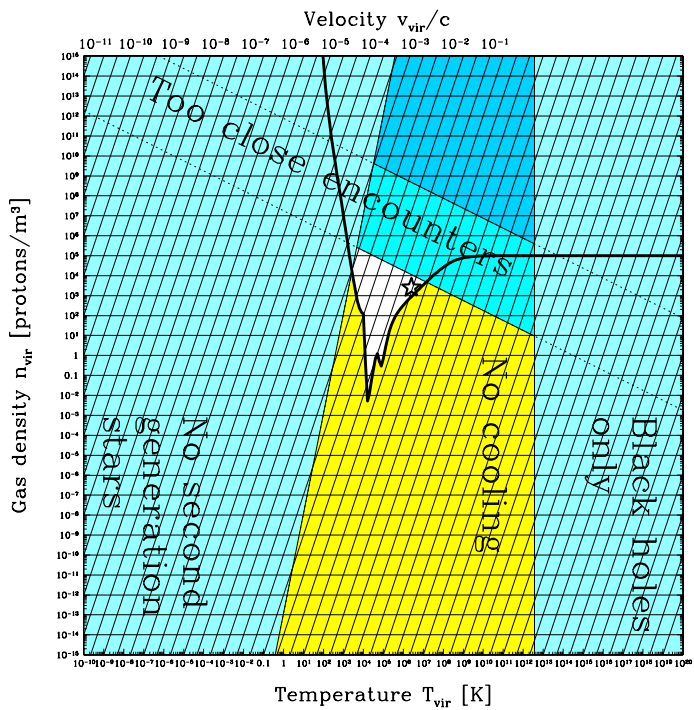}}
\caption[1]{\label{Tn5Fig}\footnotesize%
Many selection effects that we discuss are conveniently summarized in the
plane tracking the virial temperatures and densities
in dark matter halos.
The gas cooling requirement prevents halos below the heavy black curve
from forming galaxies. Close encounters make stable solar systems
unlikely above the downward-sloping line. Other dangers include collapse into black holes and disruption of galaxies by supernova explosions.
The location and shape of the 
small remaining region (unshaded) is independent 
of all cosmological parameters except the baryon fraction.
}
\end{figure}

\begin{figure} 
\centerline{\epsfxsize=\figsize\epsffile{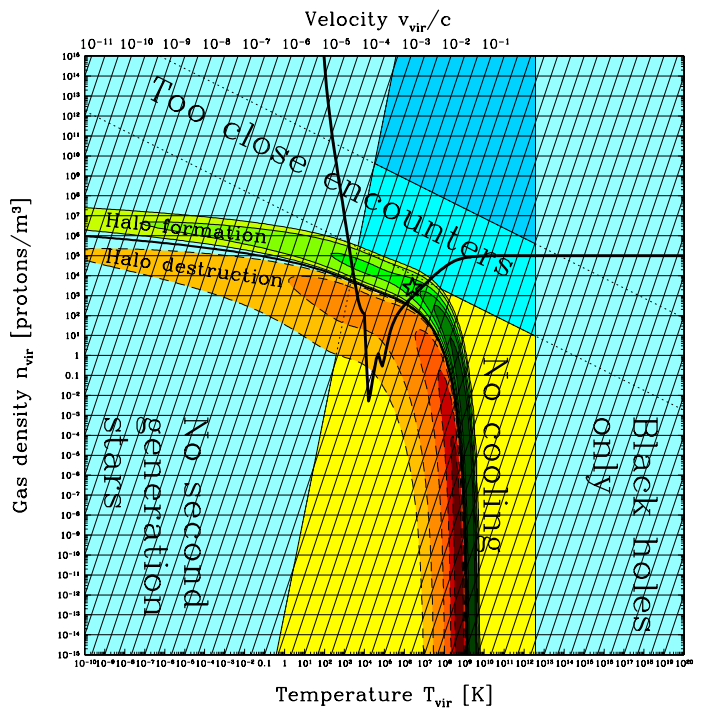}}
\caption[1]{\label{TnFig}\footnotesize%
Same as the previous figure, but including the
banana-shaped contours showing the
the halo formation/destruction rate. 
Solid/greenish contours correspond to positive net rates (halo production) at 0.5, 0.3, 0,2, 0.1, 0.03 and 0.01 of maximum,
whereas dashed/reddish contours correspond to negative net rates (halo destruction from merging) at -0.3, -0,2, -0.1, -0.03 and -0.01 
of maximum, respectively. The heavy black contour corresponds to zero net formation rate.
The cosmological parameters $(\xib,\xic,\xin,,Q,\rhol)$ have a strong effect on this banana: 
$\xib$ and $\xic$ shift this banana-shape vertically, $Q$ shifts if along the parallel diagonal lines and
$\rhol$ cuts it off below $16\rhol$ (see the next figure). 
Roughly speaking, there are stable habitable planetary systems only for cosmological parameters 
where a greenish part of the banana falls within the allowed white region from \fig{Tn5Fig}.
}
\end{figure}

We now consider selection effects, by choosing our ``selection object" to be a stable solar system, and focusing on requirements for creating these.
In line with the preceding discussion, our main interest will be to explore constraints in the 4-dimensional 
cosmological parameter space 
$(\xib,\xic,\rhol,Q)$. Since we can only plot one or two dimensions at a time, our discussion will
be summarized by a table (Table~{\ConstraintTable}) and a series of figures 
showing various 1- and 2-dimensional 
projections: $(\xib,\xic)$, $(\rhol,Q\xi)$, $Q^3\xi^4$, $\xi$, $\rhol/Q^3\xi^4$, $(\xi,Q)$.

Many of the the physical effects that lead to these constraints
are summarized in \fig{Tn5Fig} and \fig{TnFig}, showing temperatures and densities of galactic halos.
The constraints in this plane from galaxy formation and solar system stability depend only on 
the microphysical parameters $(\mp,\alpha,\beta)$ and sometimes on the baryon fraction $\xib/\xic$, 
whereas the banana-shaped constraints from dark matter halo formation 
depend on the cosmological parameters $(\xib,\xic,\xin,\rhol,Q)$, so combining them constrains certain parameter combinations.
Crudely speaking, $\fselec(\p)$ will be non-negligible only if the cosmological parameters are such that 
part of the ``banana'' falls within the ``observer-friendly" (unshaded) region sandwiched between the 
galaxy formation and solar system stability constraints

In the following three subsections, 
we will now discuss the three above-mentioned levels of structure formation in turn: halo formation, 
galaxy formation and solar-system stability.

\begin{figure} 
\centerline{\hglue-0.85cm\epsfxsize=9.3cm\epsffile{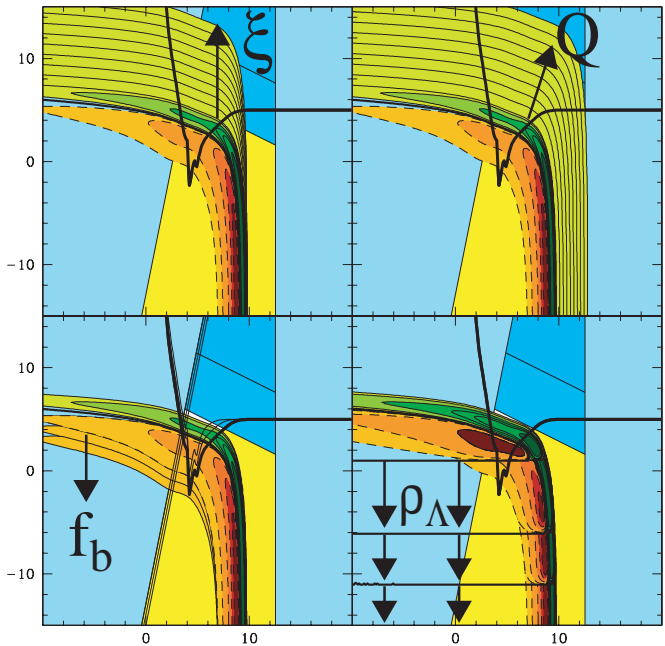}}
\caption[1]{\label{ScalingFig}\footnotesize%
Same as \fig{TnFig}, but showing effect of changing the cosmological parameters.
Increasing the matter density parameter $\xi$ shifts the banana up by a factor $\xi^4$ (top left panel), 
increasing the fluctuation level $Q$ shifts the banana up by a factor $Q^3$ and to the 
right by a factor $Q$ (top right panel), 
increasing the baryon fraction $\fb=\xib/\xi$ shifts the banana up by a factor $\fb$ 
and affects the cooling and 2nd generation constraints (bottom left panel),
and increasing the dark energy density $\rhol$ bites off the banana below $\rhovir=16\rhol$ (bottom right panel).
In these these four panels, the parameters have been scaled relative to their observed values as follows:
up by $1/4$, $2/4$,...,$9/4$ orders of magnitude ($\xi$), 
up by $1/3$, $2/3$,...,$9/3$ orders of magnitude ($Q$), 
down by $1$ and $2$ orders of magnitude ($\fb$), 
and down by $0$, $7$, $12$ and $\infty$ orders of magnitude ($\rhol$; other panels have $\rhol=0$).
}
\end{figure}

\subsection{Halo formation and the distribution of halo properties}
\label{HaloSec}

Previous studies (\eg, {\cite{BarrowTipler,LindeLambda,Weinberg87,Efstathiou95,Vilenkin95,Martel98,GarrigaVilenkin03,inflation}) 
have computed the total mass fraction collapsed into halos, as a function of cosmological parameters. 
Here, however, we wish to apply selection effects based on halo properties such as density and temperature, 
and will compute the formation rate of halos 
(the banana-shaped function illustrated in \fig{TnFig}) in terms of dimensionless parameters alone.

\subsubsection{The dependence on mass and time}

As previously explained, we will focus on the dark-matter-dominated case $\xic\gg\xib$, $\xic\gg\xin$, so
we ignore massive neutrinos and have $\xi=\xib+\xic\sim\xic$.
For our calculations, it is convenient to 
define a new dimensionless time variable
\beq{xDefEq}
x\equiv{\rhol\over\rhom}={\Omega_\Lambda\over\Omega_m}={\Omega_m^{-1}-1}
\eeq
and a new dimensionless mass variable 
\beq{muDefEq}
\mu\equiv\xi^2 M.
\eeq
In terms of the usual cosmological scale factor $a$, 
our new time variable therefore scales as $x\propto a^3$. It equals unity at the vacuum domination epoch
when linear fluctuation growth 
grinds to a halt. 
The horizon mass at matter-radiation equality is of order $\xi^{-2}$ \cite{Q}, so $\mu$ can be interpreted
as the mass relative to this scale.  It is a key physical scale in our problem.   It marks the well-known break in the matter power spectrum; 
fluctuation modes on smaller scales entered the horizon during radiation domination, when they could not grow.

We estimate the fraction of matter collapsed into dark matter halos
of mass $M\ge\xi^{-2}\mu$ by time $x$ using 
the standard Press-Schechter formalism \cite{PressSchechter}, which gives
\beq{PSeq1}
F(\mu,x)=\erfc\left[{\delta_c(x)\over\sqrt{2}\sigma(\mu,x)}\right].
\eeq
Here $\sigma(\mu,x)$ is the {\rms} fluctuation amplitude at time $x$ in 
a sphere containing mass $M=\xi^{-2}\mu$,
so $F$ is the probability that a fluctuation lies $\delta_c$ standard deviations
out in the tail of a Gaussian distribution.
As shown in \App{sFitSec}, $\sigma$ is well approximated as\footnote{Although 
the baryon density affects $\sigma$ mainly via the sum $\xi=\xib+\xic$, there is a slight correction because 
fluctuations in the
baryon component do not grow between matter-radiation equality and the drag epoch 
shortly after recombination \cite{EisensteinHu99}.
Since the resulting correction to the fluctuation growth factor ($\sim 15\%$ for the observed baryon fraction
$\xib/\xic\sim 1/6$) is negligible for our purposes, we ignore it here.
}
\beq{sigmaEq}
\sigma(\mu,x)\approx \sigmastar {Q\xi^{4/3}\over\rhol^{1/3}} s(\mu)\Gl(x),                  
\eeq
where 
\beq{sigmastarEq}
\sigmastar\equiv {45\cdot 2^{1/3}\zeta(3)^{4/3}\over\pi^{14/3}\left[1+{21\over 8}\left({4\over 11}\right)^{4/3}\right]}
\approx 0.206271
\eeq
and the known dimensionless functions $s(\mu)$ and $\Gl(x)$ do not depend on any physical parameters.
$s(\mu)$ and $\Gl(x)$ give the dependence on scale and time, respectively,
and appear in equations\eqn{sEq} and\eqn{GlambdaFitEq} in Appendix A.
The scale dependence is $s(\mu)\sim\mu^{-1/3}$ on large scales $\mu\gg 1$, saturating to only logarithmic 
growth towards small scales for $\mu\ll 1$.
Fluctuations grow as $\Gl(x)\approx x^{1/3}\propto a$ 
for $x\ll 1$ and then asymptote to a constant amplitude corresponding to 
$\Ginf\equiv 5\Gamma\left({2\over 3}\right)\Gamma\left({5\over 6}\right)/3\sqrt{\pi}
\approx 1.43728$ as $a\to\infty$ and dark energy dominates.
(This is all for $\rhol > 0$; we will treat $\rhol<0$ in \Sec{BananaScalingSec} and find
that our results are roughly independent of the sign of $\rhol$, so that we can sensibly
replace $\rhol$ by $|\rhol|$.)

Returning to \eq{PSeq1},
the collapse density threshold $\delta_c(x)$ 
is defined as the linear perturbation theory overdensity that a top-hat-averaged fluctuation would have had at the time $x$ when it collapses.
It was computed numerically in \cite{anthroneutrino}, and found to 
vary only very weakly (by about 3\%) with time,
dropping from the familiar cold dark matter value 
$\delta_c(0)=(3/20)(12\pi)^{2/3}\approx 1.68647$ early on
to the limit 
$\delta_c(\infty)= (9/5) 2^{-2/3} \Ginf\approx 1.62978$ \cite{Weinberg87}
in the infinite future.
Here we simply approximate it by the latter value:
\beq{deltacChoiceEq}
\delta_c(x)\approx\delta_c(\infty)
= {9\Gamma\left({2\over 3}\right)\Gamma\left({5\over 6}\right)\over 3\pi^{1/2}2^{2/3}} \approx 1.62978.
\eeq
Substituting equations\eqn{sigmaEq} and\eqn{deltacChoiceEq} into \eq{PSeq1}, we thus obtain
the collapsed fraction
\beq{PSeq2}
F(\mu,x)=\erfc\left[{A\rhol^{1/3}\over \xi^{4/3}Q\Gl(x)s(\mu)}\right],
\eeq
where 
\beq{AdefEq}
A\equiv {\delta_c(\infty)\over\sqrt{2}\sigma_*}
={\left(1+{42\over 22^{4/3}}\right)\pi^{25\over 6}\Gamma\left({2\over 3}\right)\Gamma\left({5\over 6}\right)
  \over 30\sqrt{2}\zeta(3)^{4/3}}
\approx 5.58694.
\eeq
Below we will occasionally find it useful to rewrite \eq{PSeq2} as 
\beq{PSeq3}
F(\mu,x)=\erfc\left[{A(\rhol/\rhostar)^{1/3}\over\Gl(x)s(\mu)}\right],\quad\rhostar\equiv\xi^4 Q^3.
\eeq

Let us build some intuition for \eq{PSeq2}.
It tells us that early on when $x\sim 0$, 
no halos have formed ($F\approx 0$),
and that as time passes and $x$ increases, small halos form before any large ones since
$s(\mu)$ is a decreasing function. Moreover, we see that since 
$\Gl(x)$ and $s(\mu)$ are at most of order unity, 
no halos will ever form if $\rhol\gg\xi^4 Q^3$.
We recognize the combination $\rhostar\equiv\xi^4 Q^3$ as the characteristic density of the universe
when halos would form in the absence of dark energy \cite{Q}.  If $\rhol\gg\rhostar$, then dark energy dominates
long before this epoch and fluctuations never go nonlinear.
\Fig{halodensityFig} illustrates that the density distributions corresponding to \eq{PSeq3} are 
broadly peaked around $\rhovir\sim 10^2\rhostar$ when $\rhostar\gg\rhol$,
are exponentially suppressed for $\rhostar\ll\rhol$, and in all cases give no halos with $\rhovir<16\rhol$.

\subsubsection{The dependence on temperature and density}

\begin{figure} 
\centerline{\epsfxsize=\figsize\epsffile{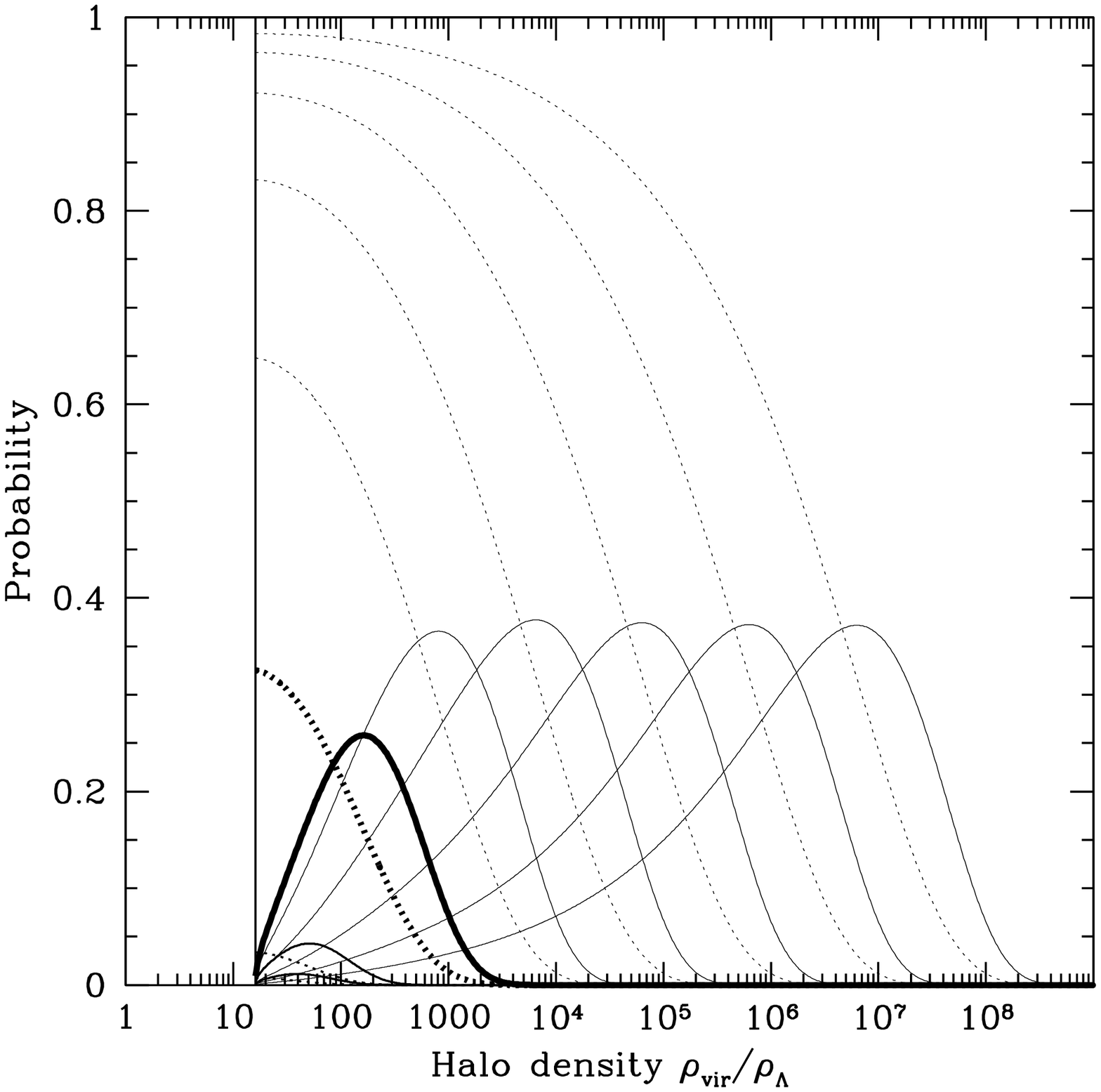}}
\caption[1]{\label{halodensityFig}\footnotesize%
Halo density distribution for various values of $\rhostarbar/\rhol$, where $\rhostarbar\equiv[s(\mu)/A]^3\rhostar\sim Q^3\xi^4$.
The dashed curves show the cumulative distribution 
$F = \erfc\left[{(\rhol/\rhostarbar)^{1/3}/\Gl(x[\rhovir])}\right]$ and the solid curves show the
probability distribution $-\partial F/\partial\lg\rhovir$ for (peaking from right to left)
$\rhostarbar/\rhol=10^5$, $10^4$, $10^3$, $10^2$, $10$, $1$ (heavy curves), $0.1$ and $0.05$. 
Note that no halos with $\rhovir<16\rhol$ are formed. 
}
\end{figure}

We now discuss how our halo mass $\mu$ and formation time $x$ transform into the astrophysically relevant 
parameters $(\Tvir,\nvir)$ appearing in \fig{TnFig}.

As shown in \App{rhovirFitSec}, halos that virialize at time $x$
have a characteristic density of order 
\beq{rhovirEq}
\rhovir\sim 16\rhol\left[\left({9\pi^2\over 8 x}\right)^{107\over 200}+1\right]^{200\over 107},
\eeq
\ie, essentially the larger of the two terms $16\rhol$ and $18\pi^2\rhom(x)$.
For a halo of total (baryonic and dark matter) mass $M$ in Planck units, this corresponds to a characteristic
size $R\sim (M/\rhovir)^{1/3}$,
velocity $\vvir\sim (M/R)^{1/2}\sim (M^2\rhovir)^{1/6}$ 
and virial temperature $\Tvir\sim m_p v_{vir}^2\sim\mp M^{2/3}\rhovir^{1/3}$, so
\beq{TvirEq}
\Tvir\sim m_p \left({16\rhol\mu^2\over\xi^4}\right)^{1/3} \left[\left({9\pi^2\over 8 x}\right)^{107\over 200} + 1\right]^{200\over 321}.
\eeq
Inverting equations\eqn{rhovirEq} and\eqn{TvirEq} gives
\beqa{muEq}
\mu 	&\sim&\sqrt{\xi^4\Tvir^3\over m_p^3\rhovir},\\
x	&\sim&{9\pi^2\over 8}\left[\left({\rhovir\over 16\rhol}\right)^{107\over 200}-1\right]^{-{200\over 107}}.\label{xEq}
\eeqa
For our applications, the initial gas temperature will be negligible and the gas density will trace
the dark matter density, so until cooling becomes important (\Sec{CoolingSec}), the proton number density
is simply 
\beq{nvirEq}
\nvir = {\xib\over\xi\mp}\rhovir.
\eeq

According to the Press-Schechter approximation, the derivative $-{\partial F\over\partial\lg\mu}(\mu,x)$ can be interpreted as the
so-called {\it mass function}, \ie, as the distribution of halo masses at time $x$.
We can therefore interpret the second derivative 
\beq{mxBananaEq}
\fmx(\mu,x)\equiv -{\partial^2F\over\partial(\lg\mu)\partial(\lg x)}
\eeq
as the net formation rate of halos as a function of mass and time.
Transforming this rate from $(\lg\mu,\lg x)$-space to $(\lg\Tvir,\lg\rhovir)$-space,
we obtain the function whose banana-shaped contours are shown in \fig{TnFig}:
\beq{TrBananaEq}
f(\Tvir,\nvir) \equiv |J| \fmx(\mu,x) = |J| {\partial^2F\over\partial(\lg\mu)\partial(\lg x)},
\eeq
where $J$ is a Jacobian determinant  that can be ignored safely.\footnote{It makes sense to treat 
$\fmx$ as a distribution, since $\int\int\fmx d(\ln\mu)\>d(\ln x)$ equals the total collapsed fraction.
When transforming it, we therefore factor in
the Jacobian of the transformation from $(\lg\mu,\lg x)$ to $(\lg\Tvir,\lg\rhovir)$,
\beq{JacobianEq}
\J\equiv
\left(\begin{tabular}{cc}
${\partial\lg\mu\over\partial\lg T}$	&${\partial\lg\mu\over\partial\lg\rho}$\\
${\partial\lg x\over\partial\lg T}$	&${\partial\lg x\over\partial\lg\rho}$,
\end{tabular}\right) 
= 
\left(\begin{tabular}{cc}
${3\over 2}$	&$-{1\over 2}$\\
$0$		&$-\left[1-\left({16\rhol\over\rhovir}\right)^{159\over 200}\right]^{-1}$
\end{tabular}\right),
\eeq
with determinant 
\beq{JacobianEq2}
J\equiv |\J| = -{3\over 2}\left[1-\left({16\rhol\over\rhovir}\right)^{159\over 200}\right]^{-1}
= -{3\over 2}\left({\rhovir x\over 18\pi^2\rhol}\right)^{159\over 200}.
\eeq
So the Jacobian is an irrelevant constant $J=-3/2$ for $\rhovir\gg 16\rhol$.
Since the entire function $f$ vanishes for $\rhovir<16\rhol$, 
the Jacobian only matters near the $\rhovir=16\rhol$ boundary, where it has a rather
unimportant effect (the divergence is integrable). \Eq{JacobianEq} shows that away from that boundary, 
the $(\lg\Tvir,\lg\rhovir)$-banana is simply a linear transformation of the 
$(\lg\mu,\lg x)$-banana $\partial^2F/\partial\lg\mu\partial\lg x$, 
with slanting parallel lines of slope 3 in \fig{TnFig} corresponding to constant $\mu$-values.
}

Let us now build some intuition for this important function $f(\Tvir,\nvir)$.
First of all, since \eq{rhovirEq} shows that $\rhovir\propto\nvir$ decreases with 
time and is $\mu$-independent, we can reinterpret the vertical $\nvir$-axis in 
\fig{TnFig} as simply the time axis: as our Universe expands, halos can form at lower densities
further and further down in the plot. Since nothing ever forms with $\rhovir<16\rhol$, the 
horizontal line $\nvir=16\rhol\xib/\xi\mp$ corresponds to $t\to\infty$.
Second, $f$ is the {\it net} formation rate, which means that it is negative
if the rate of formation of new halos of this mass is smaller than 
the rate of destruction from merging into larger halos.
The destruction stems from the fact that, to avoid double-counting, 
the Press-Schechter approximation
counts a given proton as belonging to at most one halo at a given time, 
defined as the largest nonlinear structure that it is part of.
\footnote{Our treatment could be improved by modeling halo substructure survival,
since subhalos that harbor stable solar systems
may be counted as part of an (apparently inhospitable) larger halo.
}
\Fig{TnFig} shows that halos of any given mass (defined by the lines of slope 3)
are typically destroyed in this fashion some time after its formation unless 
$\rhol$-domination terminates the process of fluctuation growth.

We will work out the detailed dependence of this banana on physical parameters 
in the next section.  For now, 
we merely note that \fig{TnFig} shows that the first halos form with characteristic
density $\rhovir\sim\rhostar\sim\xi^4 Q^3$, 
$\nvir\sim\xib\xi^3 Q^3/\mp$, 
and (assuming $\rhol\ll\rhostar$) the largest 
halos approach virial velocities $\vvir\sim Q^{1/2}c$ and 
temperatures $\Tvir\sim Q\times\mp c^2$ \cite{Q}.

\subsubsection{How $\mp$, $\xib$, $\xic$, $Q$ and $\rhol$ affect ``the banana''}
\label{BananaScalingSec}

Substituting the preceding equations into \eq{TrBananaEq} gives the explicit expression
\beqa{TrBananaEq2}
&f(\Tvir,\nvir;\mp,\xib,\xic,Q,\rhol)=\nonumber\\
&= {3a\left[\Gl(x)^2 s(\mu)^2-2 a^2\right] x\Gl'(x) \mu s'(\mu)\over \sqrt{\pi}\Gl(x)^4 s(\mu)^4\exp\left[\left({a\over\Gl(x)s(\mu)}\right)^2\right]} 
\left({\rhovir x\over 18\pi^2\rhol}\right)^{159\over 200},\nonumber\\
&a\equiv {A\rhol^{1/3}\over Q\xi^{4/3}},
\eeqa
where $\mu$ and $\x$ are determined by $\Tvir$ and $\nvir$ via equations\eqn{muEq}, \eqn{xEq} and \eqn{nvirEq}. 
This is not very illuminating.
We will now see how this complicated-looking function $f$ of seven variables can be well approximated and understood 
as a fixed banana-shaped function of merely two variables, which gets translated around by variation of 
$\mp$, $\xib$, $\xic$ and $Q$, and truncated from below at a value determined by $\rhol$.

Let us first consider the limit $\rhol\to 0$ where dark energy is negligible.
Then $x\ll 1$ so that $\Gl(x)=x^{1/3}=(\rhol/\rhom)^{1/3}$, causing \eq{PSeq2} to simplify to
\beq{PS_norholeq}
F(\mu,x)=\erfc\left[{A\rhom^{1/3}\over \xi^{4/3}Q s(\mu)}\right].
\eeq
In the same limit, \eq{rhovirEq} reduces to $\rhovir=18\pi^2\rhom$, 
so using \eq{nvirEq} gives $\rhom=\rhovir/18\pi^2=\xi\mp\nvir/18\pi^2\xib$.
Using this and \eq{muEq}, we can rewrite \eq{PS_norholeq} in the form
\beqa{PS_norholeq2}
F&=&\erfc\left[{A\left({\mp\nvir\over\xib\xi^3 Q^3}\right)^{1/3}\over\left(18\pi^2\right)^{1/3}s\left[\left({\Tvir\over\mp Q}\right)^{3/2}\left({\mp\nvir\over\xib\xi^3 Q^3}\right)^{-1/2}\right]}\right]\nonumber\\
 &=&B\left({\mp\nvir\over\xib\xi^3 Q^3},{\Tvir\over\mp Q}\right),
\eeqa
where the ``standard banana'' is characterized by a function of only two variables,
\beq{StandardBananaEq}
B(X,Y) = \erfc\left[{AX^{1/3}\over\left(18\pi^2\right)^{1/3}s\left[Y^{3/2}X^{-1/2}\right]}\right].
\eeq
Thus $B(X,Y)$ determines the banana shape, and the parameters $\mp$, $\xib$, $\xi$ and $Q$ merely shift it
on a log-log plot: 
increasing $\mp$ shifts it down and to the right along lines of slope $-1$, 
increasing $\xib$ shifts it upward, 
increasing $\xi$ shifts it upward three times faster, and 
increasing $Q$ shifts it upward and to the right along lines of slope $3$.
The halo formation rate defined by its derivatives (equation \ref{TrBananaEq}) and plotted in 
\fig{TnFig} clearly scales in exactly the same way with these four parameters, since
in the $\rhol\to 0$ limit that we are considering, the Jacobian $\J$ is simply a constant matrix.

Let us now turn to the general case of arbitrary $\rhol$.
As as discussed above, time runs downward in \fig{TnFig} since the cosmic expansion gradually 
dilutes the matter density $\rhom$. The matter density completely dwarfs 
the dark energy density at very early times.
They key point is that since $\rhol$ has no effect until our Universe has expanded enough 
for the matter density to drop near the dark energy density, the part of the banana in \fig{TnFig} 
that lies well above the vacuum density $\rhol$ will be completely independent of $\rhol$.
As the matter density $\rhom$ drops below $\rhol$ (\ie, as $x$ grows past unity), 
fluctuation growth gradually stops. This translates into a firm cutoff below 
$\rhovir=16\rhol$ in \fig{TnFig} (\ie, below $\nvir=16\xib\rhol/\mp\xi$), since \eq{rhovirEq} shows that 
no halos ever form with densities below that value.

Viewed at sensible resolution on our logarithmic plot, spanning many orders of magnitude in density, 
the transition from weakly perturbing the banana to biting it off is quite abrupt, 
occurring as the density changes by a factor of a few.
For the purposes of this paper, 
it is appropriate to approximate the effect of $\rhol$ as simply 
truncating the banana below the cutoff density.

Putting it together, we can approximate the non-intuitive \eq{TrBananaEq2} by the much simpler
\beqa{BananaApproxEq}
&f(\Tvir,\nvir;\mp,\xib,\xic,Q,\rhol)\approx\nonumber\\
&\approx b\left({\Tvir\over m_p Q},{\mp\nvir\over\xib\xi^3 Q^3}\right)\theta\left(\nvir-{16\xib\rhol\over\mp\xi}\right),
\eeqa
where $\theta$ is the Heaviside step function ($\theta(x)=0$ for $x<0$, $\theta(x)=1$ for $x\ge 0$)
and $b$ is the differentiated ``standard banana'' function 
\beq{StandardBananaEq2}
b(X,Y)\equiv{\partial^2\over\lg X\lg Y} B(X,Y). 
\eeq
This is useful for understanding constraints on the cosmological parameters:
the seemingly complicated dependence of the halo distribution on seven variables
from \eq{TrBananaEq2} can be intuitively understood as the standard banana shape from \fig{TnFig}
being rigidly translated by the four parameters $\mp$, $\xib$, $\xi$ and $Q$ 
and truncated from below  with a cutoff $\nvir>16\xib\rhol/\mp\xi$.
All this is illustrated in \fig{ScalingFig}, which also confirms numerically that the approximation of
\eq{StandardBananaEq2} is quite accurate.

\subsubsection{The case $\rhol<0$}

Above we assumed that $\rhol\ge 0$, but for the purposes of this paper, we can 
obtain a useful approximate generalization of the results by simply replacing $\rhol$ by  $|\rhol|$ in 
\eq{BananaApproxEq}.
This is because $\rhol$ has a negligible effect early on when $|\rhol|\ll\rho_m$
and a strongly detrimental selection effect when $|\rhol|\sim\rho_m$, either by
suppressing galaxy formation (for $\rhol>0$) or by recollapsing our universe (for $\rhol<0$), 
in either case causing a rather sharp lower cutoff of the banana.
As pointed out by Weinberg \cite{Weinberg87}, one preferentially expects $\rhol>0$ as observed because the constraints 
tend to be slightly stronger for negative $\rhol$.
If $\rhol>0$, observers have time to evolve long after $|\rhol|=\rhom$ 
as long as galaxies had time to form before while $\rhol$ was still subdominant.
If $\rhol<0$, however, both galaxy formation {\it and} observer evolution must be completed before $\rhol$ dominates
and recollapses the universe; thus, for example, increasing $Q$ so as to make structure form earlier will not significantly improve prospects for observers.

\subsection{Galaxy formation}
\label{GalaxySec}

\subsubsection{Cooling and disk formation}
\label{CoolingSec}

Above we derived the time-dependent (or equivalently, density-dependent) fraction of matter collapsed into halos above a given mass (or
temperature), derived from this the formation rate of halos of a given temperature and density, and discussed how both functions
depended on cosmological parameters.  
For the gas in such a halo to be able to contract and form a galaxy, it must be able to dissipate energy by cooling
\cite{Binney77,ReesOstriker77,WhiteRees78,Blanchard92}} --- see \cite{RipamontiAbel05} for a recent review.
In the shaded region marked ``No cooling'' in \fig{Tn5Fig}, the cooling timescale $T/\dot T$ exceeds 
the Hubble timescale $H^{-1}$.\footnote{This criterion is closely linked to the question of whether observers can form 
if one is prepared to wait an {\it arbitrarily} long time. 
First of all, if the halo fails to contract substantially in a Hubble time,
it is likely to lose its identity by being merged into a larger halo on that timescale unless $x\gg 1$ so that $\rhol$ has frozen clustering growth. 
Second, if $\Tvir\simlt 10^4$K so that the gas is largely neutral,
then the cooling timescale will typically be much longer than the timescale on which baryons evaporate from the halo,
and once less than $0.08\Ms$ of baryons remain, star formation is impossible.
Specifically, for a typical halo profile, about $1\%$ of the baryons in the high tail of the Bolzmann velocity distribution exceed the 
halo escape velocity, and the halo therefore loses this fraction of its mass each relaxation time (when this high tail is repopulated by collisions
between baryons). In contrast, the cooling timescale is linked to how often such collisions lead to photon emission.
This question deserves more work to clarify whether all $\Tvir\gg 10^4$K halos would cool and form stars {\it eventually} (providing their protons do not have time to decay).
However, as we will see below in \Sec{ResultsSec}, this question is unimportant for the present paper's prime focus on axion dark matter, since once we marginalize over $\rhol$, it is rather the {\it upper} 
limit on density in \fig{TnFig} that affects our result.
}
We have computed this familiar cooling curve 
as in \cite{Q} with updated molecular cooling from Tom Abel's code based on \cite{GalliPalla98}, which is 
available at \url{http://www.tomabel.com/PGas/}.
From left to right, the processes dominating the cooling curve are
molecular hydrogen cooling (for $T\simlt 10^4$K), 
hydrogen line cooling (1st trough), 
helium line cooling (2nd trough), 
{\it Bremsstrahlung} from free electrons (for $T\simgt 10^5$K)
and Compton cooling against cosmic microwave background photons (horizontal line for $T\simgt 10^9$K).
The curve corresponds to zero metallicity, since we are interested in whether the {\it first} galaxies can form.

The cooling physics of course depends only on atomic processes, \ie, on the three parameters
$(\alpha,\beta,\mp)$. 
Requiring the cooling timescale to not exceed some fixed timescale would therefore give a curve independent of all
cosmological parameters. 
Since we are instead requiring the cooling timescale 
($\propto n^{-1}$ for all processes involving particle-particle collisions) not to exceed the Hubble timescale
($\propto \rhom^{-1/2}\propto(n/\fb)^{-1/2}$), our cooling curve will depend also on the baryon fraction $f_b$, 
with all parts except the Compton piece to the right scaling vertically as $\fb^{-1}$.

\subsubsection{Disk fragmentation and star formation}
\label{FragmentationSec}

We have now discussed how fundamental parameters determine whether dark matter halos form and whether gas in such halos can cool efficiently.
If both of these conditions are met, the gas will radiate away kinetic energy and settle into a rotationally supported disk, 
and the next question becomes whether this disk is stable or will fragment and form stars. As described below, this depends strongly on the 
baryon fraction $\fb\equiv\xib/\xi$, observed to be $\fb\sim 1/6$ in our universe. The constraints from 
this requirement propagate directly into \fig{xiqFig} in \Sec{SummarySec} rather than \fig{TnFig}.

First of all, stars need to have a mass of at least $\Mmin\approx 0.08\Ms$ for their core to be hot enough to allow fusion. 
Requiring the formation of at least one star in a halo of mass $M$ therefore gives the constraint
\beq{EnoughBaryonsEq}
\fb > {\Mmin\over M}. 
\eeq
An interesting point  \cite{HellermanWalcher05} is that this constraint places an upper bound on the dark matter density parameter.
Since the horizon mass at equality is $\Meq\sim\xi^{-2}$ \cite{Q},
the baryon mass within the horizon at equality is $\Meq\xib/\xi\sim\xib/\xi^3$, which drops with increasing $\xi$.
This scale corresponds to the bend in the banana of \Fig{TnFig}, so unless \cite{HellermanWalcher05}
\beq{EnoughBaryonsEq2}
\fb \simgt {\Mmin\xi^2}, 
\eeq
one needs to wait until long after the first wave of halo formation to form the first halo containing enough gas to make a star, 
which in turn requires a correspondingly small $\rhol$-value.
The ultimate conservative limit is requiring that $\Mbh\sim\rhob H^{-3}$, the baryon mass within the horizon, exceeds $\Mmin$. 
Since $\Mbh$ increases during matter domination and decreases during vacuum domination, taking its maximum value $\Mbh\sim\fb\rhol^{-1/2}$
when $x\sim 1$,
the $\Mbh>\Mmin$ requirement gives
\beq{EnoughBaryonsEq3}
\fb \simgt {\Mmin\rhol^{1/2}}. 
\eeq
However, these upper limits on the dark matter parameter density parameter $\xic$ are very weak: 
for the observed values of $\xib$ and $\rhol$ from Table~\FundParTable, 
equations\eqn{EnoughBaryonsEq2} 
and\eqn{EnoughBaryonsEq3} give 
$\xic<\xi\simlt (\xib/\Mmin)^{1/3}\sim 10^{-22}$
and
$\xic<\xi\simlt \xib/\Mmin\rhol^{1/2}\sim 10^{-4}$,
respectively, 
limits many orders of magnitude above the observed value  $\xic\sim 10^{-28}$.

It seems likely, however, that  $\Mmin$ is a gross underestimate of the baryon requirement for star formation.
The fragmentation instability condition for a baryonic disc is essentially that it should be
self-gravitating in the ``vertical'' direction. This is equivalent to
requiring the baryonic density in the disc exceed that of the 
dark matter background it is immersed in. 
The first unstable mode is then the
one that induces  breakup into spheres of radius of order the disc thickness. 
This conservative criterion is weaker than the classic Toomre instability criterion \cite{Toomre64},
which requires the disc to be self-gravitating in the {\it radial} direction and leads to spiral arm formation.
If the halo were a singular
isothermal sphere, then \cite{FallEfstathiou80} instability would
require 
\beq{fbEq1}
\fb\simgt\lambda\left({\Tmin\over\Tvir}\right)^{1/2},
\eeq
Here $\Tmin$ is the minimum temperature that the gas can cool to
and $\lambda$ is the dimensionless specific angular momentum parameter, 
which has an approximately lognormal distribution centered around 0.08 \cite{FallEfstathiou80}.
For an upper limit on the dark matter parameter $\xic$, what matters is thus
the upper limit on $\lambda$, the upper limit on $\Tvir$ and the lower limit on $\Tmin$,
all three of which are quite firm.
Ignoring probability distributions, 
taking $\Tmin\sim\alpha^2\beta\mp/6\ln\alpha^{-1}\sim 10^4\K\sim 1\ \eV$
(atomic Hydrogen line cooling) and $\Tvir\sim 500 \eV$ (Milky Way) gives 
$\Tvir/\Tmin\sim 500$, $\xic\simlt 300\xib$ from \eq{fbEq1}.
Taking the extreme values
$\Tmin\sim 500\K$ ($H_2$-cooling freezeout) and $\Tvir\sim\mp Q\sim 20$ keV (largest clusters) gives 
$\Tvir/\Tmin\sim 10^{10}Q$ and $\xi\simlt 10^6 Q^{1/2}\xib\sim 10^4 \xib$.
On the other extreme, if we argue that $\Tvir\sim\Tmin$ for the very first galaxies to form, then 
\eq{fbEq1} gives $\xi/\xib\simlt 12$, which is interestingly close to what we observe.

For NFW potentials \cite{NFW}, $\lambda$-dependence is more complicated and the local
velocity dispersion of the dark matter near the center is lower than the
mean $\Tvir$.

Even if the baryon fraction were below the threshold of \eq{fbEq1}, stars could eventually
form because viscosity (even just ``molecular'' viscosity) would
redistribute mass and angular momentum so that the gas becomes more
centrally condensed. 
The condition then becomes that the mass of spherical blob of radius $R\sim M_{\rm d}/\Tmin$ exceed the dark mass $M_{\rm d}$ within that radius.
For the
isothermal sphere, this would automatically happen, but for a realistic
flat-bottomed or NFW potential, then this gives a non-trivial
inequality. For instance, in a parabolic potential well, the requirement
would be that
\beq{fbEq2}
\fb\simgt\left({\Tmin\over\Tvir}\right)^{3/2}.
\eeq
Although a weaker limit limit than \eq{fbEq1},
giving $\xic\simlt 10^4\xib$ for the above example with $\Tvir/\Tmin\sim 500$,
it is still stronger than those of \eq{EnoughBaryonsEq2} and \eq{EnoughBaryonsEq3}.
A detailed treatment of these issues is beyond the scope of the present paper; it should include
modeling of the $t\to\infty$ limit as well as merger-induced star formation and 
the effect of dark matter substructure disturbing and thickening the disk.

\subsection{Second generation star formation}
\label{2ndGenerationSec}

Suppose that all the above conditions have been met so that a halo has formed where gas has cooled and produced at least one star.
The next question becomes whether the heavy elements produced by the death of the first star(s) can be recycled into a solar system
around a second generation star, thereby allowing planets and perhaps observers made of elements other than
hydrogen, helium and the trace amounts of deuterium and lithium left over from big bang nucleosynthesis.

The first supernova explosion in the halo will release not only heavy metals, but also heat energy of order
\beq{SNeq}
E = 10^{-3}\mp^{-2}\sim 10^{46}\hbox{J} = 10^{51}\hbox{erg}.
\eeq
Here $\mp^{-2}$ is the approximate binding energy of a Chandrashekar mass at its Schwarzschild radius,
with the prefactor incorporating the fact that neutron stars are usually somewhat 
larger and heavier and, most importantly, that about 99\% of the binding energy is lost 
in the form of neutrinos.

By the virial theorem, the gravitational binding energy $E$ of the halo equals twice its total kinetic energy, \ie,
\beq{Eeq}
E\approx M\vvir^2\sim\sqrt{\Tvir^5\over\mp^5\rhovir},
\eeq
where we have used \eq{muEq} in the last step.
If $\Esn\gg E$, the very first supernova explosion will therefore expel essentially all the gas from the halo,
precluding the formation of second  generation stars. Combining equations\eqn{SNeq} and\eqn{Eeq}, we therefore obtain the constraint
\beq{2ndGenerationEq}
\rhovir\simlt 10^6{\Tvir^5\over\mp}.
\eeq
\Fig{Tn5Fig} illustrates this fact that lines of constant binding energy have slope 5, and 
shows that the second generation constraint rules out an interesting part of the the $(\nvir,\Tvir)$ plane that is allowed by both cooling
and disruption constraints.

While we have ignored the important effect of cooling by gas in the supernova's 
immediate environment, this constraint is probably nonetheless rather 
conservative, and a more detailed calculation may well move it further to the right.
First of all, many supernovae tend to go off in close succession in a star formation site, thereby jointly releasing more energy than 
indicated by \eq{Eeq}.
Second, it is likely that many supernovae are required to produce sufficiently high metallicity. 
Since $\sim 1$ supernova forms per $100\Ms$ of star formation, releasing $\sim 1\Ms$ of metals, raising the mean 
metallicity in the halo to solar levels ($\sim 10^{-2}$) would require an energy input of order 
$E/(100\mstar/\mp)\sim 10^{-5}\mp\sim 10\,$keV per proton.
A careful calculation of the corresponding temperature would need to model the gas 
cooling occurring between the successive supernova explosions.

\subsection{Encounters and extinctions}
\label{EncounterSec}

The effect of halo density on solar system destruction was discussed in \cite{Q} making the crude assumption
that all halos of a given density had the same characteristic velocity dispersion.
Let us now review this issue, which will play a key role in determining predictions for the axion density, from the slightly more refined perspective of \fig{Tn5Fig}.
Consider a habitable planet orbiting a star of mass $M$ suffering a close encounter with
another star of mass $\Mdeath$, approaching with a relative velocity $\vdeath$ and an
impact parameter $b$.
There is some ``kill'' cross section $\sigmadeath(M,\Mdeath,\vdeath)=\pi b^2$ corresponding to encounters
close enough to make this planet uninhabitable. 
There are several mechanisms through which this could happen:
\begin{enumerate}
\item It could become gravitationally unbound from its parent star, thereby losing its key heat source.
\item It could be kicked into a lethally eccentric orbit.
\item The passing star could cause disastrous heating.
\item The passing star could perturb an Oort cloud in the outer parts of the solar 
      system, triggering a lethal comet impact.
\end{enumerate}
The probability of the planet remaining unscathed for a time $t$ is then $e^{-\deathrate t}$,
where the destruction rate is $\deathrate$ is
$\nstar\sigmadeath\vdeath$ appropriately averaged over incident velocities $\vdeath$ and 
stellar masses $M$ and $\Mdeath$.
Assuming that $\nstar\propto\nvir$, $\vdeath\propto\vvir\propto\Tvir^{1/2}$ and $\sigmadeath$ is independent of 
$\nvir$ and $\Tvir$, contours of constant destruction rate in \fig{Tn5Fig} are thus lines of slope $-1/2$.
The question of which such contour is appropriate for our present discussion is highly uncertain,
and deserving of future work that would lie beyond the scope of the present paper. 
Below we explore only a couple of crude estimates, based on direct and indirect impacts, respectively.

\subsubsection{Direct encounters}

Lightman \cite{Lightman84} has shown that if the planetary surface temperature is
to be compatible with life as we know it, the orbit around the central
star should be fairly circular and have a radius of order
\beq{rAUeq}
\rau\sim\alpha^{-5}\mp^{-3/2}\beta^{-2}\sim 10^{11}\Meter,
\eeq
roughly our terrestrial ``astronomical unit'',
precessing one radian in its orbit on a timescale
\beq{tYearEq}
t_{orb}\sim\alpha^{-15/2}\mp^{-5/4}\beta^{-3}\sim 0.1 \hbox{ year}.
\eeq
An encounter with another star with impact parameter $r\simlt\rau$
has the potential to throw the planet into a highly eccentric orbit
or unbind it from its parent star.\footnote{Encounters have a negligible effect on our orbit
if they are adiabatic, \ie, if the impact duration 
$r/v\gg t_{orb}$ so that the solar system
returned to its unperturbed state once the encounter 
was over.
Encounters are adiabatic for
\beq{vAdiabEq}
v \simlt {r\over c\torb} \sim{\alpha^{5/2}\beta\over\mp^{1/4}}\sim 0.0001\sim 30\km/\Second,
\eeq
the typical orbital speed of a terrestrial planet.
As long as the impact parameter $\simlt \rau$, however, the encounter is guaranteed to be non-adiabatic and hence dangerous, since
the infalling star will be gravitationally accelerated to at least this speed.
}
For $\nstar$, what matters here is not the typical stellar density in a halo, but the stellar density near other stars,
including the baryon density enhancement due to disk formation and subsequent fragmentation.
Let us write
\beq{nstarfudgeEq}
\nstar={\nstarfudge\nvir\over\Nstar},
\eeq
where $\Nstar\sim\mp^{-3}\sim 10^{57}$ is the number of protons in a star and the dimensionless 
factor $\nstarfudge$ parametrizes our uncertainty about the extent to which stars concentrate near other stars.
Substituting characteristic values $\nvir\sim 10^3/m^3$ for the Milky Way and 
$\nstar\sim (1\,\pc)^{-3}$ for the solar neighborhood gives a 
concentration factor $\nstarfudge\sim 10^5$.
Using this value, $\vdeath\sim\vvir$ and $\sigmadeath=\pi\rau^2$ excludes the dark shaded region to the upper right in \fig{Tn5Fig}
if we require $\deathrate^{-1}$ to exceed 
$t_{\rm min}\sim 10^9$ years $\sim\alpha^2\alpha_g^{-3/2}\beta^{-2}$,
the lifetime of a bright star \cite{CarrRees79}.
It is of course far from clear what is an appropriate evolutionary or geological timescale to use here,
and there are many other uncertainties as well.
It is probably an overestimate to take $\vdeath\sim\vvir$ since we only care about relative velocities. 
On the other hand, our value of $\sigmadeath$ is an underestimate since we have neglected gravitational focusing.
The value we used for $\nstarfudge$ is arguably an underestimate as well, stellar densities being substantially higher
in giant molecular clouds at the star formation epoch.

\subsubsection{Indirect encounters}

In the above-mentioned encounter scenarios 1-3, the 
incident directly damages the habitability of the planet. 
In scenario 4, the effect is only indirect, sending a hail of comets towards the inner solar system which may at a later time impact
the planet. This has been argued to place potentially stronger upper limits on $Q$ than direct encounters \cite{GarrigaVilenkin05}.

Although violent impacts are commonplace in our particular solar system, large uncertainties remain in the statistical details thereof
and in the effect that changing $\nstar$ and $\vdeath$ would have.
It is widely believed that solar systems are surrounded by a rather spherical cloud of comets composed of ejected leftovers.
Our own particular Oort cloud is estimated to contain of order 
$10^{12}$ comets, extending out to about a lightyear ($\sim 10^5$AU) from the Sun.
Recent estimates suggest that the impact rate of Oort cloud comets exceeding 1 km in diameter 
is between  5 and 700 per million years \cite{Nurmi01}. 
These impacts are triggered by gravitational perturbations to the Oort cloud 
sending a small fraction of the comets into the inner solar system.
About 90\% of these perturbations are estimated to be caused by Galactic tidal forces 
(mainly related to the motion of the Sun with respect to the Galactic midplane), with random passing stars being 
responsible for most of the remainder and random passing molecular clouds playing a relatively minor role
\cite{Byl83,Hut97,Nurmi01}.

It is well-known that Earth has suffered numerous violent impacts with celestial bodies in the the past, and the 1994 impact of
comet Shoemaker-Levy on Jupiter illustrated the effect of comet impacts.
Although the nearly 10 km wide asteroid that hit the Yucatan 65 million years ago \cite{Alvarez80}
may actually have
helped our own evolution by eliminating dinosaurs, a larger impact of a 30 km object 
3.47 billion years ago \cite{Byerly02} 
may have caused a global tsunami and massive heating,
killing essentially all life on Earth
(For comparison, the Shoemaker-Levy fragments were less than 2 km in size.)
We therefore cannot dismiss out of hand the possibility that we are in fact close to the edge in parameter space, with
only a modest increase in comet impact rates on planets causing a significant drop in the fraction of planets evolving 
observers. This possibility is indicated by the light-shaded excluded region in the upper right of \fig{Tn5Fig}.

However, there are large uncertainties here of two types.
First, we lack accurate risk statistics for our own particular solar system.
Although the lunar crater radius distribution is roughly power law of slope $-2$ at high end,
we still have very limited knowledge of the size distribution of comet nuclei \cite{Meech04} 
and hence cannot accurately estimate the frequency of extremely massive impacts.
Second, we lack accurate estimates of what would happen in denser galaxies.
There, the more frequent close  encounters  with other stars could rapidly strip stars of much of their dangerous Oort cloud, so it is far from
obvious that the risk rises as $\nstar\vdeath$. One interesting possibility is 
that the inner Oort cloud at radii $\simlt 1000$AU
contributes a substantial fraction of our impact risk, in which case such tidal stripping of most of the cloud by volume 
will do little to reduce risk.

An indirect hint that comet impacts are anthropically important may be the observation \cite{Gonzalez99}
that the orbit of our Sun through the galaxy appears fine-tuned to minimize Oort cloud perturbations and resulting comet impacts: 
compared to similar stars, its orbit has an unusually low eccentricity and small amplitude of vertical motion relative to the Galactic disk.

\subsubsection{Nearby explosions}

A final category of risk that deserves further exploration is that from nearby supernova explosions and gamma-ray bursts.
Since these are independent of stellar motion and thus depend only on $\nstar$, not on $\Tvir$, they would correspond to a horizontal upper
cutoff in \fig{Tn5Fig}.
For example, the Ordovician extinction 440 million years ago has been blamed on a nearby gamma-ray burst.
This would have depleted the ozone layer causing a massive increase in ultraviolet solar radiation exposure
and could also have triggered an ice age \cite{Melott05}.

\subsection{Black hole formation}
\label{BHsec}
There are two potentially rather extreme selection effects involving black holes.

First, \fig{Tn5Fig} illustrates a vertical constraint on the right side corresponding to 
large $\Tvir$-values.
Since the right edge of the halo banana is at $\vvir\sim Q^{1/2}c$, 
typical halos will form black holes for $Q$-values of order unity as discussed in \cite{Q}.
Specifically, 
typical fluctuations would be of black-hole magnitude already by the time
they entered the horizon, converting some substantial fraction of the
radiation energy into black holes shortly after the end of inflation and continually increasing this
fraction as longer wavelength fluctuations entered and collapsed.
For a scale-invariant spectrum, 
extremely rare fluctuations that are $Q^{-1}$ standard deviations out in 
the Gaussian tail can cause black hole domination if \cite{Q}
\beq{QbhEq}
Q\simgt 10^{-1}.
\eeq
This constraint is illustrated in \fig{xiqFig} below rather than in \fig{TnFig}.

A second potential hazard occurs if halos form with high enough density that collapsing gas can trap photons.
This makes the effective $\gamma$-factor close to 4/3 so that the 
Jeans mass does not fall as collapse proceeds, and collapse proceeds in a qualitatively different way because there will be no tendency to fragment.  This might
lead to production of single black hole (instead of a myriad of stars) or other pathological objects, but what actually happens would depend on unknown details, such as whether angular momentum can be transported outward so as to prevent the formation of a disk.

\subsection{Constraints related to the baryon density}

\begin{figure} 
\centerline{\epsfxsize=\figsize\epsffile{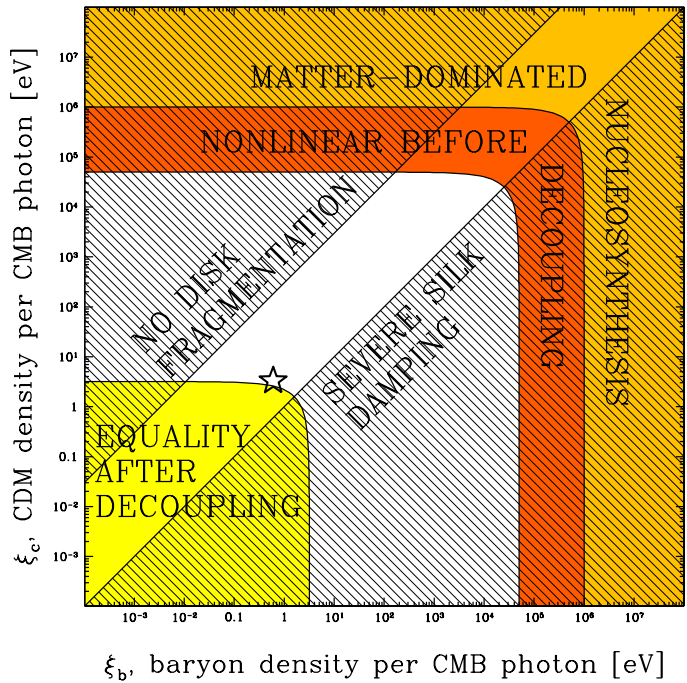}}
\caption[1]{\label{xibxicFig}\footnotesize%
Qualitatively different parts of the $(\xib,\xic)$-plane.
Our observed universe corresponds to the star, and altering the parameters to leave the white region corresponds to qualitative 
changes. The hatched regions delimit baryon fractions $fb=\xib/\xic>300$ and $fb<1$, which may suppress 
galaxy formation by impeding disk fragmentation and by Silk damping out Galaxy-scale matter clustering, respectively,
The shaded regions delimit interesting ranges of constant total matter density parameter $\xi=\xib+\xic$ --- to what
extent these qualitative transitions are detrimental to galaxy formation deserves further work.
}
\end{figure}

To conclude our discussion of selection effects,
\fig{xibxicFig} illustrates a number of constraints that are independent of $Q$ and $\rhol$, 
depending on the density parameters $\xib$ and $\xic$ for baryonic and dark matter
either through their ratio or their sum $\xi\equiv\xib+\xic$.

As we saw in \Sec{FragmentationSec}, a very low baryon fraction $\fb\equiv\xib/\xi\simlt 300$
may preclude disk fragmentation and star formation.
On the other hand, a baryon fraction of order unity corresponds to dramatically suppressed matter clustering on
Galactic scales, since Silk damping around the recombination epoch suppresses fluctuations in 
the baryon component and the fluctuations
that created the Milky Way were preserved through this epoch mainly by dark matter \cite{cmbfast}.

Recombination occurs at $T\sim $Ry$/50\sim\mp\alpha^2\beta/100\sim 3000\K$, whereas matter-domination 
occurs at $\Teq\sim 0.2\xi$.
If $\xi\simlt 0.05\mp\alpha^2\beta$, recombination would therefore precede matter-radiation equality,
occurring during the radiation-dominated epoch. It is not obvious whether this would have detrimental effects
on galaxy formation, but it is interesting to note that, as illustrated in \fig{xibxicFig},
our universe is quite close to this boundary in parameter space.

If we instead increase $\xi$, a there are two qualitative transitions.

In the limit $x=\rhol/\rhom\ll 1$, using \eq{sigmaEq}, \eq{ngEq} and 
$\rhom=\xi\ng$ gives
\beq{TnonlinearEq}
\sigma = \left({\pi^2\over 2\zeta(3)}\right)^{1/3}\sigmastar s(\mu){\xi Q\over T}\approx 0.33028\,s(\mu){\xi Q\over T}.
\eeq 
This means that the first Galactic mass structures ($s(\mu)\approx 28$) go nonlinear at $T\sim 9 \xi Q$, \ie, before recombination if 
$\xi\simgt 10^{-3}\mp\alpha^2\beta\mp/Q$.
A baryonic cloud able to collapse before or shortly after recombination while the 
ionization fraction remained substantial would trap radiation and, 
as mentioned above in \Sec{BHsec}, potentially produce a single black hole instead of stars.

Big Bang Nucleosynthesis (BBN) occurs at $T\sim\mp\beta_n\sim 1\MeV$, so 
if we further increase $\xi$ so that $\xi\simgt\mp\beta_n$, then
the universe would be matter-dominated before BBN, producing dramatic (but not necessarily fatal \cite{Aguirre99})
changes in primordial element abundances.

\section{Putting it all together}
\label{ResultsSec}

As discussed in the introduction, theoretical predictions for physical parameters are confronted with observation using 
\eq{BayesEq}, where neither of the two factors $\fprior(\p)$ and $\fselec(\p)$ are optional.
Above we discussed the two factors for the particular case study of cosmology and dark matter, 
covering the prior term in \Sec{PriorSec} and the selection effect term in \Sec{SelectionSec}.
Let us now combine the two and investigate the implications for the parameters $\xi$, $Q$ and $\rhol$.

\begin{figure} 
\centerline{\epsfxsize=\figsize\epsffile{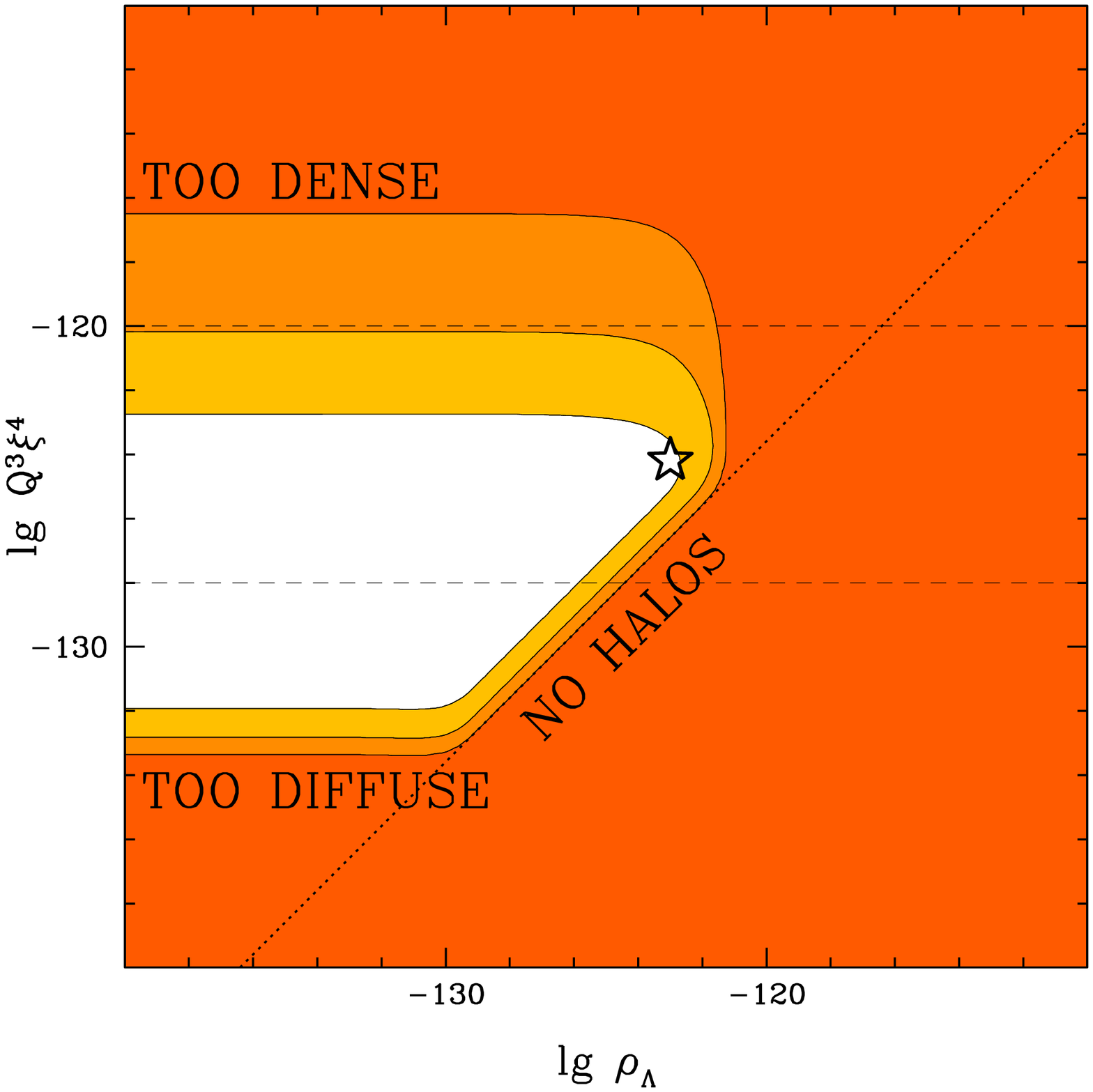}}
\caption[1]{\label{rholrhostarFig}\footnotesize%
The fraction of all protons that end up in halos with density in the range
$\rhomin<\rhovir<\rhomax$ is shown as a function of the cosmological parameters
$\rhol$ and $\rhostar\equiv Q^3\xi^4$. The contours show fractions $0.3$, $0.05$ and $0.003$
for the example 
$\rhomin=10^{-128}$, $\rhomax=10^{-120}$ (the two dashed lines) and 
$\mu=10^{-5}$ (corresponding roughly to $10^{12}\Ms$ halos and giving $s(\mu)\approx 28$).
When $\rhol\gg Q^3\xi^4$, essentially no halos at all are formed.
When $Q^3\xi^4\simlt 10^{-3}\rhomin$, essentially all halos have $\rhovir<\rhomin$.
When $Q^3\xi^4\simgt 10^2\rhomax$, essentially all halos have $\rhovir>\rhomax$.
The star shows our observed parameter values.
}
\end{figure}

\subsection{An instructive approximation}

For this exercise, we would ideally want to compute, say, the function $\fselec(\p)$
defined as the fraction of protons ending up in stable habitable planets as function of $(\rhol,\xi,\xib,Q)$,
leaving the particle physics parameters fixed at their observed values.
\Fig{TnFig} shows that this is quite complicated, for two reasons. First, as discussed above we have 
computed only certain integrated versions of $\fselec$. Second,
the other selection effects discussed involve
substantial uncertainties.
To provide useful qualitative intuition, let us therefore start by working out the implications of the simple approximation
that there are sharp upper and lower halo density cutoffs, \ie, that observers only form 
in halos within some density range $\rhomin\le\rhovir\le\rhomax$, where these two density limits may depend on $Q$, $\xi$ and $\xib$. 
Based on the empirical observation that typical galaxies lie near the right side of the cooling curve in \Fig{Tn5Fig},
one would expect $\rhomin$ to be determined by the bottom of the cooling curve and $\rhomax$ by
the intersection of the cooling curve with the encounter constraint.

\Eq{PSeq3} showed that
the fraction of all protons collapsing into a halo of mass $\mu$
is $\erfc[A(\rhol/\rhostar)^{1/3}/s(\mu)\Ginf]$, where 
$\rhostar\equiv Q^3\xi^4$ can be crudely interpreted as the characteristic density of the first
halos to form.
Roughly, $A/s(\mu)\Ginf\sim 1$, $\xi^4$ is the matter-radiation equality density
and $Q^3$ is the factor by which our Universe gets diluted between equality (when fluctuations start to grow)
and galaxy formation.
More generally, of these protons, the fraction $f_h$ in halos within a density range $\rhomin\le\rhovir\le\rhomax$ is
\beq{rhominmaxLeq}
f_h(\rhol,\xi,\xib,Q) = \erfc\left[{(\rhol/\rhostarbar)^{1/3}\over\Gl(x(\rhomin))}\right] - \erfc\left[{(\rhol/\rhostarbar)^{1/3}\over\Gl(x(\rhomax))}\right],
\eeq
where the function $x(\rho)$ is given by \eq{xEq} extended so that
$x(\rho)=\infty$ for $\rho\le 16\rhol$.
For convenience, we have here defined the rescaled density
\beq{rhostarbarEq}
\rhostarbar\equiv \left[{s(\mu)\over A}\right]^3\rhostar.
\eeq
\Fig{rholrhostarFig} is a contour plot of this function, and illustrates that in addition to
classical constraint $\rhol\simlt Q^3\xi^4$ dating back to Weinberg and others
\cite{BarrowTipler,LindeLambda,Weinberg87,Efstathiou95,Vilenkin95,Martel98,GarrigaVilenkin03},
we now have 
two new constraints as well: $\rhol$-independent upper and lower bounds on $\rhostar\equiv Q^3\xi^4$.

\subsection{Marginalizing over $\rhol$}

As mentioned in \Sec{PriorSec}, there is good reason to expect the prior $\fprior(\rhol,\xi,Q)$
to be independent of $\rhol$ across the tiny relevant range $|\rhol|\simlt Q^3\xi^4$ where $\fselec(\rhol,\xi,Q)$ is non-negligible.
This implies that we can write the
predicted  parameter probability distribution from \eq{BayesEq} as
\beq{fEq1}
f(\rhol,\xi,\xib,Q)\propto\fprior(\xi,\xib,Q) f_o(\xi,\xib,Q) f_h(\rhol,\xi,\xib,Q),
\eeq
where $f_o(\xi,\xib,Q)$ is the product of all other selection effects that we have discussed --- 
we saw that these were all independent of $\rhol$.

\begin{figure} 
\centerline{\epsfxsize=\figsize\epsffile{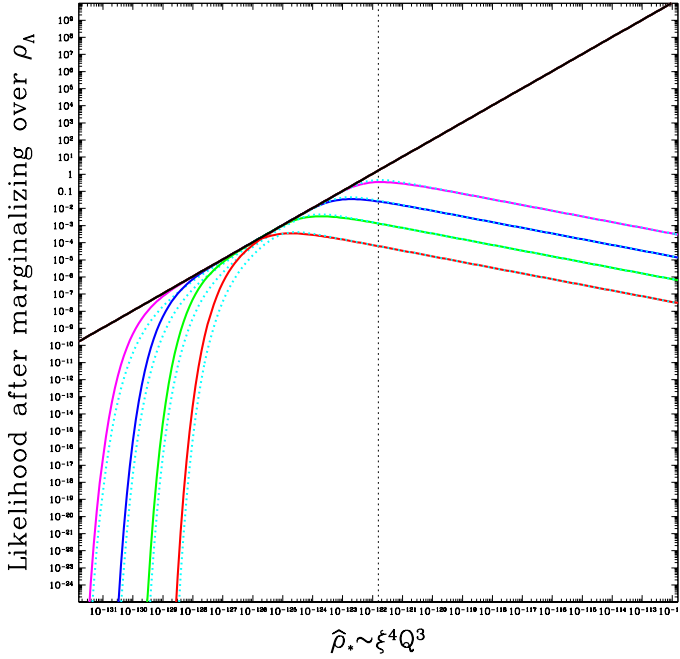}}
\caption[1]{\label{rhostarprobFig}\footnotesize%
The probability factor $f_h(\rhostar)$ after marginalizing over $\rhol$.
The  black line of  slope $1$ is for $\rhomin=0$, $\rhomax=\infty$.
Increasing $\rhomin$ shifts the exponential cutoff on the left side progressively further the right: 
$\rhomin=10^{-5}$, $10^{-4}$, $10^{-3}$ and $10^{-2}$ times the density at the dotted horizontal line 
gives the four curves on the left.
Decreasing $\rhomax$ shifts left the break to a $-1/3$ slope further to the right: 
$\rhomax=100$, $10$, $1$ and $0.1$ times the density at the dotted horizontal line 
gives the four downward-sloping curves on the right, respectively.
}
\end{figure}

The key point here is that the {\it only} $\rhol$-dependence in \eq{fEq1} comes from
the $f_h$-term, \ie, from selection effects involving the halo banana.
It is therefore interesting to marginalize over $\rhol$ and study the resulting predictions for $Q$ and $\xi$.
Thus integrating \eq{fEq1} over $\rhol$, we obtain the theoretically predicted probability distribution
\beqa{fEq2}
&&f(\xi,Q)\propto\fprior(\xi,\xib,Q) f_o(\xi,\xib,Q) f_h(\xi,\xib,Q),\quad\hbox{where}\nonumber\\
&&f_h(\xi,\xib,Q)\equiv\int_0^\infty f_h(\rhol,\xi,\xib,Q)d\rhol.
\eeqa
Note that if $\rhomin$ and $\rhomax$ are constants, then $f_h(\xi,Q)$ is really a function of only one variable:
since \eq{rhominmaxLeq} shows that $\xi$ and $Q$ enter in $f_h(\rhol,\xi,Q)$ 
only in the combination $\rhostar\equiv Q^3\xi^4$,
the marginalized result $f_h(\xi,Q)$ will be a function of $\rhostar$ alone.
\Fig{rhostarprobFig} shows this function evaluated numerically for various values of 
$\rhomin$ and $\rhomax$, as well as the following approximation (dotted curves) which becomes 
quite accurate for $\rho\ll\rhomin$ and $\rho\gg\rhomax$:
\beq{rhominmaxLmargEq}
f_h\approx
\left\{724.9{\rhostarbar^{1/3}\over\rhomax^{4/3}}  +
{\sqrt{\pi}\over \Ginf^3\rhostarbar\erfc\left[\left({\rhomin\over 18\pi^2\rhostarbar}\right)^{1/3}\right]}
\right\}^{-1}\eeq
This approximation is valid also when $\rhomin$ and $\rhomax$ are functions of the cosmological parameters,
as long as they are independent of $\rhol$.
To understand this result qualitatively, consider first the simple case 
where we count baryons in all halos regardless of density, \ie, 
$\rhomin=0$, $\rhomax=\infty$.
Then a straightforward change of variables in \eq{fEq1} gives 
$f_h(\rhostar)\propto\rhostar=Q^3\xi^4$, as was previously derived in \cite{inflation}.
This result troubled the authors of both \cite{inflation} and \cite{GarrigaVilenkin05}, since any 
weak prior on $Q$ from inflation would be readily overpowered by the $Q^3$-factor, leading to the
incorrect prediction of a much larger CMB fluctuation amplitude than observered.
The same result would also spell doom for the axion dark matter model discussed in \Sec{AxionSec}, since
the $\xi^{-1/2}$-prior would be overpowered by the $\xi^4$-factor from the selection effect and dramatically overpredict
the dark matter abundance.

\Fig{rhostarprobFig} shows that the presence of selection effects on halo density has the potential to solve this problem.
Both the problem and its potential resolution can be intuitively understood from \fig{rholrhostarFig}.
Note that $f_h(\rhostar)$ is simply the horizontal integral of this two-dimensional distribution.
For $\rhomin\ll\rhostar\ll\rhomax$, the integrand $\approx 1$ out to the $\rhostar\sim\rhol$-line 
marked ``NO HALOS'', giving  $f_h(\rhostar)\propto\rhostar$.
In the limit $\rhostar\gg\rhomax$, however, we are integrating across the region marked `TOO DENSE'' in 
\fig{rhostarprobFig},
and the result drops as $f_h(\rhostar)\propto\rhomax^{4/3}/\rhostar^{1/3}=\rhomax^{4/3}/Q\xi^{4/3}$.
This is why \fig{rhostarprobFig} shows 
a break in slope from $+1$ to $-1/3$ at a location 
$\rhostar\sim\rhomax/200$, 
as illustrated by the four curves with
different $\rhomax$-values.
Conversely, in the limit $\rhostar\ll\rhomin$, we are integrating across the exponentially suppressed 
region marked `TOO DIFFUSE'' in \fig{rholrhostarFig}, causing a corresponding exponential suppression in 
\fig{rholrhostarFig} for $\rhostar\ll\rhomin$.

\begin{figure} 
\centerline{\epsfxsize=\figsize\epsffile{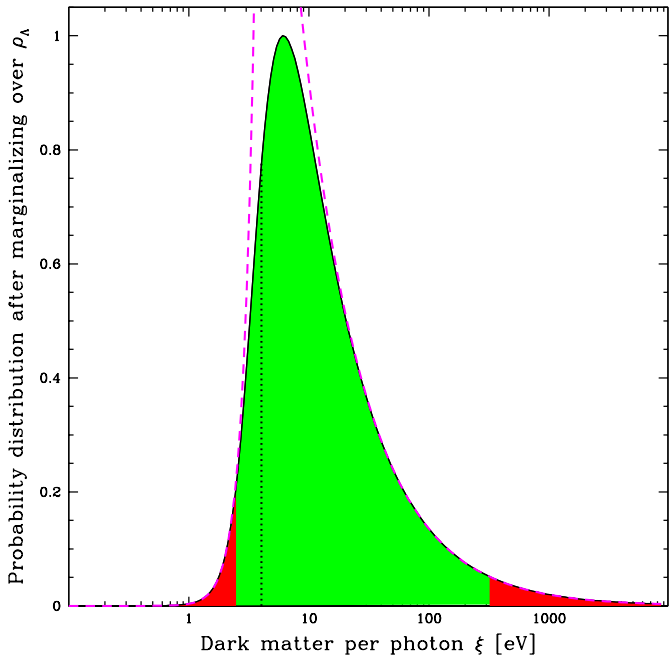}}
\caption[1]{\label{xiFig}\footnotesize%
Probability distribution for the axion dark matter density parameter $\xi$ 
measured from a random $10^{12}\Ms$ halo with virial density below 5000 times 
the present cosmic matter density.
Green/light shading indicates the 95\% confidence interval.
The dotted vertical line shows our observed value $\xi\approx 4$ eV, in good agreement with the
prediction. The dashed curves show the analytic asymptotics $\xi^{9/2}$ and 
$\xi^{-5/6}$, respectively.
}
\end{figure}
 
\begin{figure} 
\centerline{\epsfxsize=\figsize\epsffile{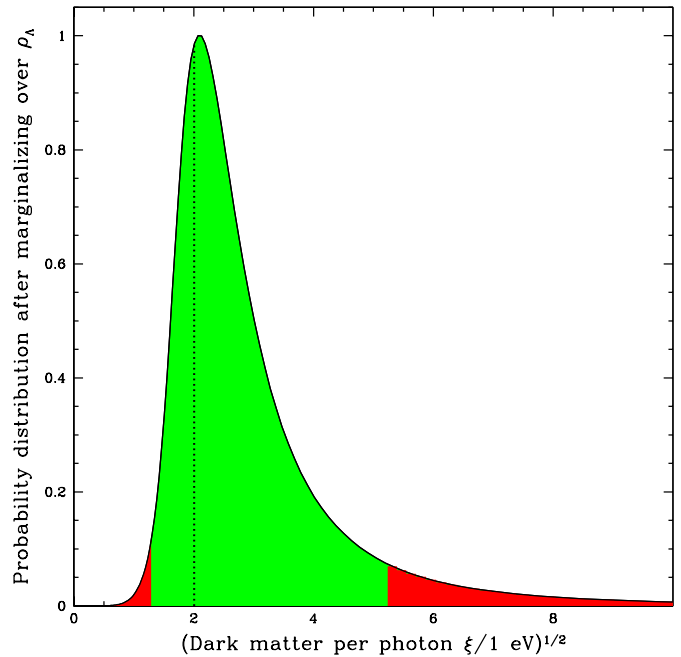}}
\caption[1]{\label{xiFig2}\footnotesize%
Same as previous figure, but showing the 
probability distribution for $\xi^{1/2}$, the quantity which has a uniform
prior in our axion model. The shape of this curve therefore reflects the selection effect alone.
}
\end{figure}
 
Figures~\ref{xiFig} and~\ref{xiFig2} 
illustrate this for the simple example of
using the \Sec{AxionSec} axion prior $\fprior(\xi)\propto\xi^{-1/2}$,
treating $Q$ as fixed at its observed value, 
ignoring additional selection effects ($f_o(\xi)=1$), and imposing a 
halo density cutoff $\rhomax$ of 5000 times the current cosmic matter density.

The key features of this plot follow directly from our analytic approximation of \eq{rhominmaxLmargEq}.
First, for $\rhostar\ll\rhomax$, we saw that $f_h\propto\rhostar\propto\xi^4$, so
\eq{fEq2} gives the probability growing as $f\propto\xi^{-1/2}\xi^4\xi=\xi^{9/2}$
(the last $\xi$-factor comes from the fact that we are plotting 
the probability distribution for $\log\xi$ rather than $\xi$).
This means that imposing a lower density cutoff $\rhomin$ would have essentially no effect. 
This also justifies our approximation of using the total matter density parameter $\xi$ as a proxy for 
the dark matter density parameter $\xic$: we have a baryon fraction $\xib/\xic\ll 1$ in the interesting regime,
with a negligible probability for $\xic\simlt\xib\sim 0.6\eV$
(the probability curve continues to drop still further to the left even without the $f_h$-factor).
Second, for $\rhostar\gg\rhomax$, we saw that $f_h\propto\rhostar^{-1/3}\propto\xi^{-4/3}$, so
\eq{fEq2} gives the probability falling as $f(\ln\xi)\propto\xi^{-1/2}\xi^{-4/3}\xi=\xi^{-5/6}$.

Although this simple example was helpful for building intuition, a more accurate treatment is required 
before definitive conclusions about the viability of the axion dark matter model can be drawn. 
One important issue, to which we return below in \Sec{ResultsSec}, 
is the effect of the unknown $Q$-prior, since the preceding equations only constrained a combination
of $Q$ and $\xi$.
A second important issue, to which we devote the remainder of this subsection, has to do with properly incorporating the constraints from \fig{TnFig}.
If we only consider the encounter constraint and make the unphysical simplification of ignoring the effect of halo velocity dispersion,
then our upper limit should be not on dark matter density ($\rhomax$ constant) but on baryon density 
($\rhomax\xib/\xi$ constant). Inserting this into \eq{rhominmaxLmargEq} makes the term
$\rhostarbar^{1/3}/\rhomax^{4/3}$ independent of $\xi$ as $\xi\to\infty$, replacing
the $f_h\propto Q^{-1}\xi^{-4/3}$ cutoff  
by $f_h\propto Q^{-1}\xib^{-4/3}$, which is constant if $\xib$ is.
The axion prior would then predict a curve $f(\ln\xi)\propto\xi^{1/2}$ rising without bound.
This failure, however, results from ignoring some of the physics from \Fig{TnFig}.
Considering a series of nominally more likely domains with progressively larger $\xi$, 
they typically have more dark energy ($\rhol\propto\xi$) and higher characteristic halo baryon density ($\nvir\propto\xi^3\xib$),
so stable solar systems are found only in rare galaxies that formed exceptionally late, just before $\rhol$-domination, 
with $\nvir\propto\rhovir\xib/\xi\simgt 16\rhol \xib/\xi\propto\xib$ roughly $\xi$-independent and below the the maximum allowed value.
Since the increased dark matter density boosts virial velocities, 
this failure mode corresponds to moving from the star in \fig{TnFig} straight to the right, running 
right into the constraints from cooling and velocity-dependent encounters.
In other words, a more careful calculation would be expected to give a prediction qualitatively similar to that 
of \fig{xiFig}. The lower cutoff would remain visually identical to that of \fig{xiFig} as long as $\rhomin$ 
can be neglected in the calculation, whereas the upper cutoff would become either steeper or shallower depending on the details of
the encounter and cooling constraints.
This interesting issue merits further work going well beyond the scope of the present paper,
extending our treatment of halo formation, halo mergers and galaxy formation into 
a quantitative probability distribution in the plane of \fig{TnFig}, \ie, into a function
$f(\Tvir,\nvir;Q,\rhol,\xi,\xib)$ that could be multiplied by the plotted constraints and marginalized over $(\Tvir,\nvir)$ to give 
$\fselec(Q,\rhol,\xi,\xib)$.

\subsection{Predictions for $\rhol$}

\begin{figure} 
\centerline{\epsfxsize=\figsize\epsffile{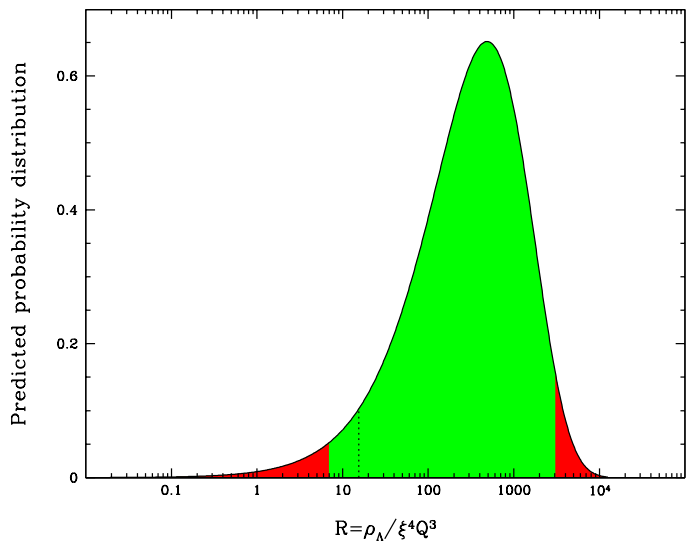}}
\caption[1]{\label{rFig}\footnotesize%
Probability distribution for the quantity $R\equiv\rhol/\xi^4 Q^3$ measured from a
random $10^{12}\Ms$ halo, using a uniform prior for $R$ and ignoring other selection effects.
This is equivalent to treating $\xi$ and $Q$ as fixed. Green/light shading indicates the 95\% confidence interval,
the dotted line indicates the observed value $R\approx 15$.
}
\end{figure}

This above results also have important implications for the dark energy density $\rhol$.
The most common way to predict its probability distribution in the literature has
been to treat all other parameters as fixed and assume both that the number of observers is proportional to 
the matter fraction in halos
(which in our notation means
$\fselec(p)=\erfc[A(\rhol/Q^3\xi^4)^{1/3}/s(\mu)\Ginf]$) and 
that
$\fprior(\rhol)$ is constant across the
narrow range $|\rhol|\simlt Q^3\xi^4$ where $\fselec$ is non-negligible.
This gives the familiar result shown in \fig{rFig}: a probability distribution
consistent with the observed $\rhol$-value but favoring slightly larger values.
The numerical origin of the predicted magnitude $\rhol\sim 10^{-123}$ is thus the {\it measured} value
of $\rhostar\equiv Q^3\xi^4$.

Generalizing this to the case where $Q$ and/or $\xi$ can vary across an inflationary multiverse, 
the predictions will depend on the precise question asked.
\Fig{rFig} then shows the successful prediction for $\rhol$ given (conditionalized on) our measured values
for $Q$ and $\xi$. However, when testing a theory, we wish to use {\it all} opportunities that we have to 
potentially falsify it, and each predicted parameter offers one such opportunity.
For our axion example, the theory predicts a 2-dimensional distribution in the
$(\rhol,\xi)$-plane of which \fig{xiFig} is the marginal distribution for $\xi$.
The corresponding marginal distribution for $\rhol$ (marginalized over $\xi$) will 
generally differ from \fig{rFig} in both its shape and in the location of its peak.
It will differ in shape because the $(\rhol,\xi)$-distribution is not generally separable: 
the selection effects (as in \fig{rFig}) will not be a function of $\rhol$ times a function of $\xi$; in other words, a uniform prior on $\rhol$ will {\it not} correspond to 
a uniform prior on the quantity $R\equiv\rhol/Q^3\xi^4$ plotted in \fig{rholrhostarFig} 
once non-separable selection effects such as ones on the halo density parameter $\rhostar\equiv Q^3\xi^4$ are included.
Moreover, the distribution will in general not peak where that in \fig{rFig} does
because the predicted magnitude $\rhol\sim 10^{-123}$
no longer comes from conditioning on astrophysical measurements of $Q$ and $\xi$, but from other parameters like 
$\alpha$, $\beta$ and $\mp$ that determine the selection effects in \fig{TnFig} and 
the maximum halo density $\rhomax$.
In this case, whether the observed $\rhol$-value agrees with predictions or not thus
depends sensitively on how strong the selection effects against dense halos are.

In summary, the prediction of $\rhol$ {\it given} $Q$ and $\xi$ is an unequivocal success,
whereas predictions for $Q$, $\xi$ and the entire joint distribution
for $(\rhol,\xi,Q)$ are fraught with the above-mentioned uncertainties.  Since a uniform $\rhol$-prior does not imply a uniform
$R\equiv\rhol/Q^3\xi^4$-prior, we cannot conclude that anthropic arguments succeed in predicting $R\equiv\rhol/Q^3\xi^4$ \cite{GarrigaVilenkin05} without additional hypotheses.

\subsection{Constraint summary}
\label{SummarySec}

\begin{figure} 
\centerline{\epsfxsize=\figsize\epsffile{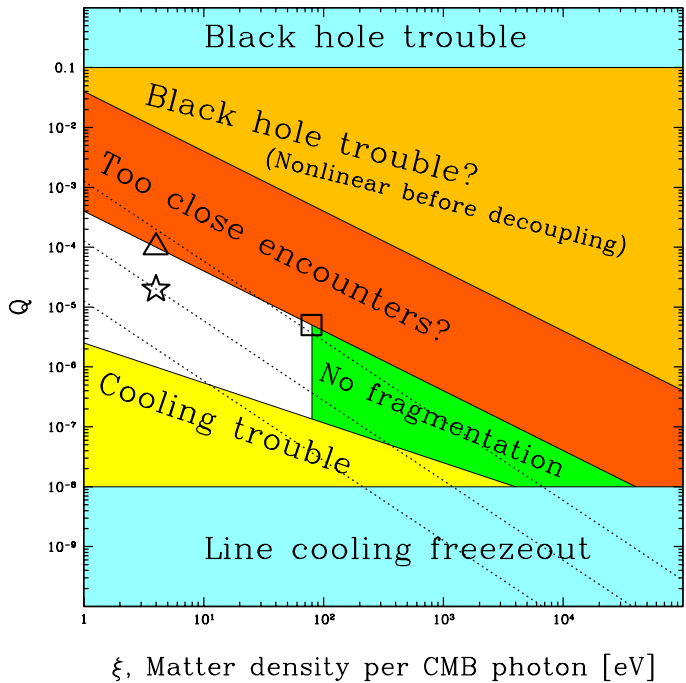}}
\caption[1]{\label{xiqFig}\footnotesize%
Crude summary of constraints $(\xi,Q)$-plane constraints from Table~{\ConstraintTable}.
The star shows our observation  $(\xi,Q)\approx (4\eV,2\times 10^{-5})$.
The parallel dotted lines are lines of constant characteristic halo density, so the constraint that
dark matter halos form (that they go nonlinear before vacuum domination freezes fluctuation growth)
rules out everything to the lower left of these lines, which from left to right correspond to
$\rhol/\rhol^{\rm obs}=10^{-3}$, $1$, $10^3$. 
This means that if $\xi$ is fixed, marginalizing over $\rhol$ pushes things to the triangle at 
{\it larger} $Q$ (the ``smoothness problem'' \cite{inflation}), but if 
$\xi$ too can vary, marginalizing over $\rhol$ instead pushes things
to the square at {\it smaller} $Q$.
}
\end{figure}

\begin{table*}
{\footnotesize 
Table~\ConstraintTable: These constraints (see text) are summarized in \fig{xiqFig}

\begin{tabular}{|l|l|l|l|}
\hline
Constraint				&Generally						&Fixing $(\alpha,\beta,\mp)$		&Fixing all but $(Q,\xi)$\\
\hline
Need nonlinear halos			&$|\rhol|\simlt\rhostar$				&$|\rhol|/\xi^4 Q^3\simlt 1$		&$Q\simgt 10^{-5}(\xi/\xi_0)^{-4/3}$\\
Avoid line cooling freezeout		&$Q\simgt\alpha^2\beta$					&$Q\simgt 10^{-8}$			&$Q\simgt 10^{-8}$\\
Primordial black hole excess		&$Q\simlt 10^{-1}$					&$Q\simlt 10^{-1}$			&$Q\simlt 10^{-1}$\\
Need cooling in Hubble time 		&$Q^3\xib^2\xi^2\simgt\alpha^{-3}\ln[\alpha^{-2}]^{-16/3}\beta^4\mp^6/125$	&$Q^3\xib^2\xi^2\simgt 10^{-129}$	&$Q\simgt 10^{-6}(\xi/\xi_0)^{-2/3}$\\
Avoid close encounters			&							&$Q^3\xib\xi^3\simlt 10^{-123}$		&$Q\simlt 10^{-4}(\xi/\xi_0)^{-1}$\\
Go nonlinear after decoupling		&$\xi Q\simlt 10^{-3}\alpha^2\beta\mp$			&$\xi Q\simlt 10^{-30}$  		&$Q\simlt 10^{-2} (\xi/\xi_0)^{-1}$\\  
Need equality before decoupling?	&$\xi\simgt 0.05\alpha^2\beta\mp$			&$\xi\simgt 10^{-28}$			&$(\xi/\xi_0)\simgt 1/3$\\
Avoid severe Silk damping		&$\fb\simlt 1/2$					&$\xi/\xib\simgt 2$			&$(\xi/\xi_0)\simgt 1/3$\\
Need disk instability 			&$\fb\simlt 10^2$					&$\xic/\xib\simlt 10^2$			&$(\xi/\xi_0)\simlt 20$\\
\hline
\end{tabular}
}
\end{table*}

Table~{\ConstraintTable}
summarizes the constraints that we have discussed above. Those not superseded by other stronger ones are 
also illustrated in \fig{xiqFig}, with $\xib$ fixed at its observed value.
Let us briefly comment on how they all fit together.

The first key point to note in \fig{xiqFig} is that the various constraints involve many different combinations of $Q$ and $\xi$,
thereby breaking each other's degeneracies and providing interesting constraints on both $\xi$ and $Q$ separately.
The fragmentation constraint (requiring disk instability) constrains $\xi$ alone when $\xib$ is given
whereas there are constraints on $Q$ alone from both line cooling freezeout and primordial black hole trouble.
The encounter and cooling constraints both have different slopes ($-1$ and $-2/3$, respectively) because they
involve different powers of the baryon fraction.
Whereas the nonlinear halo constraint depends on the total halo density $\propto Q^3\xi^4$,
the encounter constraint depends on the halo baryon density $\propto Q^3\xi^3\xib$.
The cooling constraint (which is equation (11) in \cite{Q} after changing variables to reflect the notation in this paper)
comes from equating the cooling timescale (given by the inverse of the baryon density $\propto Q^3\xi^3\xib$)
and the dynamical timescale (given by the total density as $\propto\rhovir^{-1/2}\propto (Q^3\xi^4)^{-1/2}$,
thus constraining the combination $Q^3\xi^2\xib^2$.

The second key point in \fig{xiqFig} is that this combination of constraints reverses previously published 
conclusions about $Q$.
Where we should expect to find ourselves in \fig{xiqFig} depends on the priors for both $Q$ and $\xi$.
Recall that halos form only above whichever dotted line corresponds to the $\rhol$-value.
If the prior for $\rhol$ is indeed uniform, then it is {\it a priori} (before selection effects are 
taken into account) 1000 times more likely for the halo constraint to be the upper dotted line than the middle one,
and this is in turn 1000 more likely than the bottom dotted line.
In other words, marginalizing over $\rhol$ relentlessly pushes us towards the upper right in \fig{xiqFig},
towards larger value of $\rhostar=Q^3\xi^4$.
This means that if $\xi$ is fixed and only $Q$ can vary across the ensemble, marginalizing over $\rhol$ pushes things
towards the triangle at {\it larger\/} $Q$. This ``smoothness problem'' 
was pointed out in \cite{Q} and elaborated in \cite{Garriga99,Graesser04,inflation}, suggesting that
unless the $\fprior(Q)$ falls off as $Q^{-4}$ or faster, it is overpowered by the selection effect and one
predicts a clumpier universe than observed (the star in \fig{xiqFig} should be shifted vertically up against the edge of the
encounter constraint, to the triangle).
\Fig{xiqFig} shows that if $\xi$ too can vary, marginalizing over $\rhol$ instead pushes things
towards {\it smaller\/} $Q$, to the square, since this gives the greatest allowed halo density $\propto Q^3\xi^4$.
In \cite{inflation}, it was found that certain low-energy inflation scenarios naturally
predicted a rather flat priors for $\lg Q$ --- to determine whether they are ruled out or not thus requires
more detailed modeling of the effect of the dark matter density on galaxy formation and encounters.

Table~{\ConstraintTable} illustrates that (except for one rather unimportant logarithm),
all constraints we have considered correspond to hyperplanes
in the 7-dimensional space parametrized by 
\beq{HyperspaceEq}
(\lg\alpha,\lg\beta,\lg\mp,\lg\rhol,\lg Q,\lg\xi,\lg\xib),
\eeq
providing a simple geometric interpretation of the selection effects.
This means that if all priors are power laws, the preferred parameter values 
are determined by solving a simple linear programming problem, and will generically 
correspond to one of the corners of the convex 7-dimensional allowed region.
Above we have frequently used such sharp inequalities to help build intuition for the underlying physics.
For future extensions of this work, however, it should be borne in mind
that such inequalities are not necessarily sufficient to capture the key features
of the problem, and that they can either overstate or understate the importance of 
selection effects.
The ultimate goal is to test theories by computing the predicted probability distribution from
\eq{BayesEq}, and almost no selection effects  
$\fselec(\p)$ correspond to sharp cutoffs resembling a Heaviside step function.
Some (like the halo constraint) are exponential and thus able to overpower any power law prior.
Others, however, may be softer, and it is therefore crucial to check whether their functional 
form falls of faster than than the relevant prior grows --- if not, the ``constraint'' will be penetrated,
and the most likely parameter values may lie in the allegedly ``disallowed region''. 
Conversely, if $f(\p)=\fprior(\p)\fselec(\p)$ is rather flat in a large ``allowed region'', then 
the most likely parameter values can lie well inside it, far from 
any edges, since it corresponds to the peak of the probability distribution rather than its edge.

\section{Conclusions}
\label{ConclusionsSec}

We have discussed predictions for the 31-dimensional vector $\p$ of dimensionless physical 
constants in the context of ``ensemble theories'' producing multiple Hubble volumes (``universes'') 
between which one or more of the parameters can differ.
The prediction is the probability distribution of \eq{BayesEq}, where neither of the two factors is optional.
Although it is generally difficult to evaluate either one of the terms, it is nonetheless crucial:
if candidate theories involve such ensembles, they are neither testable nor falsifiable unless their probabilistic predictions
can be computed.

Although many scientists hope that $\fprior$ in \eq{BayesEq} will be a multidimensional
Dirac $\delta$-function, rendering
selection effects irrelevant, to elevate this hope into an assumption would, ironically, be to
push the anthropic principle to a hedonistic extreme, suggesting that nature
must be devised so as to make mathematical physicists happy. We must therefore face some difficult questions both about what, in principle, to select on, and how to quantitatively calculate that selection effect.
For both statistical mechanics and quantum mechanics, it proved challenging to correctly predict 
probability distributions. If the analogous challenge can be overcome for theories predicting
dimensionless constants, they too will become testable in the same fashion. 
The fact that we can only observe once is not a show-stopper: 
a single observation of a vertically polarized photon passing through a polarizer
$\theta=89.9^\circ$ from vertical would rule out quantum mechanics at 
$1-\cos^2\theta\approx 99.9997\%$ confidence. 
The outstanding difficulty is therefore not a ``philosophical" problem that we cannot observe the full ensemble, 
but rather that we generally do not know how to compute the theoretical prediction $f(\p)$.

\subsection{Axion dark matter}

We have tackled this problem quantitatively for a particular example where both factors 
in \eq{BayesEq} {\it are} computable: that
of axion dark matter with its phase transition well before the end of inflation.
We found that $\fprior(\xi)\propto\xi^{-1/2}$, \ie, that the prior dark matter distribution has no free parameters
even though the axion model itself does.
We saw that this useful predictability gain has the same basic origin as 
that of the zero-parameter (flat) $\rhol$-prior of \cite{Weinberg87}:
the anthropically 
relevant range over which $\fselec$ is non-negligible is much smaller 
than the natural range of $\fprior$, so only a particular property of $\fprior$ matters.
For the famous $\rhol$-case, the natural $\fprior$-range involves either the supersymmetry-breaking scale or the Planck scale $|\rhol|\simlt\rhoplanck$ whereas
the relevant range is $|\rhol|\simlt 10^{-123}\rhoplanck$, so the property that $\fprior(\rhol)$ is smooth on that tiny scale implies that 
it is for all practical purposes constant.
For our $\xi$-case, the natural $\fprior$-range is $0\le\xi\le\xistar$ where 
$\xistar^{1/4}$ lies somewhere between the inflation scale and the Planck scale, whereas the relevant range is much smaller, so the only property of $\fprior(\xi)$
that matters is its asymptotic scaling as $\xi\to 0$.

In computing the selection effect factor $\fselec(\rhol,\xi,Q)$, we took advantage of the fact that 
none of these parameters affect chemistry or biology directly. We could therefore avoid poorly understood biochemical issues related to life and consciousness, and could limit our calculations
to astrophysical selection effects involving halo formation, galaxy, solar system stability, {\etc}
The same simplifying argument has been previously applied to $\rhol$ and the neutrino density (\eg, \cite{anthrolambdanu}).
We discovered that combining various astrophysical selection effects with all three parameters
$(\rhol,Q,\xi)$ as variables 
reversed previous conclusions (\fig{xiqFig}), pushing towards lower rather than higher $Q$-values.
In particular, we found that one can reach misleading conclusions by simply taking $\fselec$ to be 
the matter fraction collapsing into halos, neglecting the fact that very dense halos limit planetary stability.
In particular, we found that if we impose a stiff upper limit on the density of habitable halos, the predicted probability distribution for the dark matter density parameter (\fig{xiFig}) can be
brought into agreement with the measured $\xi$-value for reasonable assumptions, \ie, that the pre-inflationary axion model
is a viable dark matter theory.

\subsection{Multiple dark matter components}

Suppose that supersymmetry were to be discovered at the Large Hadron Collider, indicating the existence of 
a weakly interacting massive particle (WIMP) with a relic density approximately equal to 
the observed dark matter density, \ie, with $\xiwimp\sim\xic$.
There would then be a strong temptation to declare the dark matter problem solved.
Based on our results, however, that would be quite premature!

As detailed in \Sec{WimpSec}, it is not implausible that the function 
$\fprior(\xiwimp)$ to be taken as input to the astrophysics calculation 
is a sharply peaked function whose relative width is much narrower than that of $\fselec(\xic)$, perhaps 
only a few percent. This situation could arise if the fundamental theory predicted a broad prior only for 
the $\mu^2$-parameter in Table~{\FundParTable} (which controls the Higgs vacuum expectation value), since this 
may arguably be determined to within a few percent from selection effects in the nuclear physics sector, notably
stellar carbon and oxygen production.
If the pre-inflationary axion model is correct, we would then have
\beq{DarkMattersEq}
\xic=\xiwimp+\xiaxion,
\eeq
where $\fprior(\xiaxion)\propto\xiaxion^{-1/2}$ and 
$\xiwimp$ would be for all practical purposes a constant.
This gives $\fprior(\xic)\propto (\xic-\xiwimp)^{-1/2}$ for 
$\xic\ge\xiwimp$, zero otherwise, and the testable prediction 
becomes this times the astrophysical factor selection effects factor 
$\fselec(\xic)$ that we have discussed.
Repeating the calculation of \cite{conditionalization}, this implies that we should expect to find roughly 
comparable densities of WIMP and axion dark matter.

As shown in \cite{conditionalization}, removing the assumption that $\fprior(\xiwimp)$ is sharply 
peaked does not alter this qualitative conclusion.
This conclusion would also hold for any other axion-like fields that were energetically irrelevant during inflation,
even unrelated to the strong CP-problem.
More generally, string-inspired model building suggests the possibility that multiple species of ``GIMPS'' 
(gravitationally interacting massive particles that do not couple through the electromagnetic, weak or strong 
interactions) may be present.  According to recent understanding, their presence may be expected for modular inflation, though not
for brane inflation \cite{DouglasPC}. 
If such particles exist and have generic not-too-extreme priors 
(falling no faster than $\xi^{-1}$ and rising no faster than the selection effect cutoff), then our anthropic constraints
on the sum of their densities again predict a comparable density for all of species of dark matter that have a rising prior 
\cite{conditionalization}.

Even before the LHC turns on, more careful calculation of the astrophysical selection effects determining $\fselec$
could give interesting hints.  If the constraints on dense halos turn out to be rather weak so that 
the improved version of \fig{xiFig} predicts substantially more axion dark matter than observed, this will favor 
an alternative model such as WIMPs where $\fprior(\xic)$ is so narrow that $\xic$ is effectively determined non-anthropically. 
Alternatively, if \fig{xiFig} remains consistent with observation, 
then it will be crucial to follow up an initial LHC SUSY or other WIMP detection, if it occurs, with detailed measurements of the relevant parameters
to establish whether the WIMP density accounted for (which a linear collider might, in favorable cases, pinpoint within a few percent \cite{Feng05}) 
is exactly equal to the cosmologically measured dark matter density,
or whether a substantial fraction of the dark matter still remains to be accounted for.

\bigskip

The authors wish to thank Ed Bertschinger, Douglas Finkbeiner, Alan Guth, Craig Hogan, Andrei Linde and Matias Steinmetz for helpful comments, and Sam Ribnick for some earlier, exploratory calculations.
Thanks to Tom Abel for molecular cooling software.
This work was supported by NASA grant NAG5-11099,
NSF CAREER grant AST-0134999, and fellowships from the David and Lucile
Packard Foundation and the Research Corporation. The work of AA was supported in part by NSF grant AST-0507117.
The work of FW is supported in part by funds provided by
the U.S.~Department of Energy under cooperative research agreement
DE-FC02-94ER40818.

\appendix

\section{Fitting functions used}

In this Appendix, we derive a variety of fitting functions used in the paper, explicitly highlighting the
dependence of standard cosmological structure formation on the dimensionless parameters $\xi$, $\rhol$, $Q$ and $\mu$.
Since some of our approximations are quite accurate, we retain the exact numerical coefficients where appropriate.

\subsection{The fluctuation time dependence $G(x)$}
\label{GfitSec}

As shown in \cite{anthroneutrino}, the  
function 
\beq{GlambdaFitEq}
\Gl(x)\approx x^{1/3}\left[1+\left({x\over \Ginf^3}\right)^\alpha\right]^{-1/3\alpha},
\eeq
where $\alpha=159/200=0.795$ and
\beq{GmaxEq}
\Ginf\equiv {5\Gamma\left({2\over 3}\right)\Gamma\left({5\over 6}\right)\over 3\sqrt{\pi}}\approx 1.43728,
\eeq
describes how, in the absence of massive neutrinos, 
fluctuations in a $\rhol>0$ $\Lambda$CDM universe grow as the cosmic scale factor $a$ as long as dark energy is negligible 
($\Gl(x)\approx x^{1/3}\propto a\propto (1+z)^{-1}$ for $x\ll 1$) and then asymptote to 
a constant value as $t\to\infty$ and dark energy dominates ($\Gl(x)\to \Ginf$ as $x\to\infty$).

Including the effect of radiation domination early on but ignoring baryon-photon coupling, the total growth factor is 
well approximated by \cite{anthroneutrino}
\beq{Gtoteq}
G(x)\approx 1+{3\over 2}\xeq^{-1/3}\Gl(x),
\eeq
which is accurate to better than 1.5\% for all $x$ \cite{anthroneutrino}. 
In essence, fluctuations grow as $\delta\propto a\propto x^{1/3}$ between 
matter domination ($x=\xeq$) and dark energy domination ($x=1$), giving a 
net growth of $\xeq^{-1/3}$.
\Eq{Gtoteq} shows that they grow by an extra factor of 1.5 by starting slightly before 
matter domination and by an extra factor $\Gl(\infty)\approx 1.44$ by
continuing to grow slightly after dark energy domination.

Let us now simplify this result further.
From the definition of $\xi$, the matter density can be written
$\rhom=\xi\ng$, where $\ng$ is the number density of photons. The latter is 
given (in Planck units) by the standard black body formula
\beq{ngEq}
\ng={2\zeta(3)\over\pi^2} T^3,\quad \zeta(3)\approx 1.20206.
\eeq
The corresponding standard formula for the energy density is
\beq{rhogEq}
\rhog={\pi^2\over 15} T^4
\eeq						
for photons and 
\beq{rhonEq}
\rhong={21\over 8}\left({4\over 11}\right)^{4/3}\rhog\approx 0.68132\rhog
\eeq						
for three standard species of relativistic massless neutrinos.
The matter-radiation equality temperature where $\rhom=\rhog+\rhong$ is thus
\beq{TeqEq}
\Teq={30\zeta(3)\over\pi^4}{\left[1+{21\over 8}\left({4\over 11}\right)^{4/3}\right]^{-1}}\xi\approx 0.220189\xi,
\eeq
measured to be about $9400\K$ in our Hubble volume.
Since $\xeq=\rhol/\rhomeq=\rhol/\xi\ng(\Teq)$, we thus obtain
\beq{xeqEq}
\xeq 
= {\pi^{14}\over 2\cdot 30^3\zeta(3)^4}\left[1+{21\over 8}\left({4\over 11}\right)^{4/3}\right]^3 {\rhol\over\xi^4}
\approx 384.554 {\rhol\over\xi^4}
\eeq
The above relations are summarized in Tables~{\DerivedParTable} and~\DerivedQuantTable.
In all cases of relevance to the present paper, the growth 
factor $G(x)\gg 1$, allowing initial fluctuations $Q\ll 1$ to grow 
enough to go nonlinear. We can therefore drop the first term in \eq{Gtoteq},
obtaining the useful result
\beqa{Gtoteq2}
G(x)&\approx&{45\cdot 2^{1/3}\zeta(3)^{4/3}\over\pi^{14/3}}\left[1+{21\over 8}\left({4\over 11}\right)^{4/3}\right]^{-1}{\xi^{4/3}\over\rhol^{1/3}}\Gl(x)\nonumber\\
&\approx&0.206271 {\xi^{4/3}\over\rhol^{1/3}}\Gl(x).
\eeqa

\subsection{The fluctuation scale dependence $s(\mu)$}
\label{sFitSec}

The {\rms} fluctuation amplitude $\sigma$ in a sphere of radius
$R$ are given by \cite{PeeblesBook80}
\beq{sigma8eq}
\sigma^2 ={4\pi} \int_0^\infty 
\left[{\sin x-x\cos x\over x^3/3}\right]^2 P(k) {k^2 dk\over(2\pi)^3},
\eeq
where $x\equiv kR$.
We are interested in the linear regime long after the relevant fluctuation modes have entered the horizon,
when all modes grow at the same rate (assuming a cosmologically 
negligible contribution from massive neutrinos as per \cite{sdsslyaf}. 
This means that $\sigma$ can be factored as a 
product of a function of time $x$ and a function of comoving scale:
\beq{sigmaFactorizationEq}
\sigma = Q G(x) s(\mu).
\eeq
Here $Q$ is the amplitude of primordial fluctuations created by, \eg, inflation, defined
as the scalar fluctuation amplitude $\delta_H$ on the horizon.
As our measure of comoving scale, we use the dimensionless quantity
\beq{muDefEq2}
\mu\equiv\xi^2 M,
\eeq
where $M$ is the comoving mass enclosed by the above-mentioned sphere.

The horizon mass at matter-radiation equality is of order $\xi^{-2}$, so $\mu$ can be interpreted
as the mass relative to this scale. This equality horizon scale is the key physical scale in 
the problem, and the one on which the matter power spectrum has the well-known break in its slope.
(fluctuation modes on smaller scales entered the horizon before equality when they could not grow).
For reference, the measured vales 
from WMAP+SDSS give 
$\xi\approx 3\times 10^{-28}$, 
$\xi^{-2}\approx 10^{17}$ solar masses and $\mu\approx R/(60 h^{-1}\Mpc)^3$
for a sphere of comoving radius $R$.
Numerically, using \eq{sigma8eq} and employing \cite{cmbfast} to compute $P(k)$, we find that 
the scale-dependence $s(\mu)$ is approximately given by 
\beq{sEq}
s(\mu)\approx \left[(9.1\mu^{-2/3})^\beta + (50.5\lg(834 + \mu^{-1/3}) - 92)^\beta\right]^{1/\beta},
\eeq
where $\beta= -0.27$.
This assumes that the primordial fluctuations are approximately scale-invariant and that 
massive neutrinos have no major effect, as indicated by recent measurements \cite{Spergel03,sdsspars,sdsslyaf}.
For our observed $\xi$-value 
a galactic mass scale
$M=10^{12}\Ms$ thus corresponds to $\mu=\xi^2 M\approx 10^{-5}$ and $s(\mu)\approx 28$.
Again using measured parameter values, 
the common reference scale $R=8h^{-1}\Mpc$ corresponds to
$\mu\approx 0.002$ and $s(\mu)\approx 11$.

\subsection{The virial density $\rhovir(x)$}
\label{rhovirFitSec}

\begin{figure} 
\centerline{\epsfxsize=\figsize\epsffile{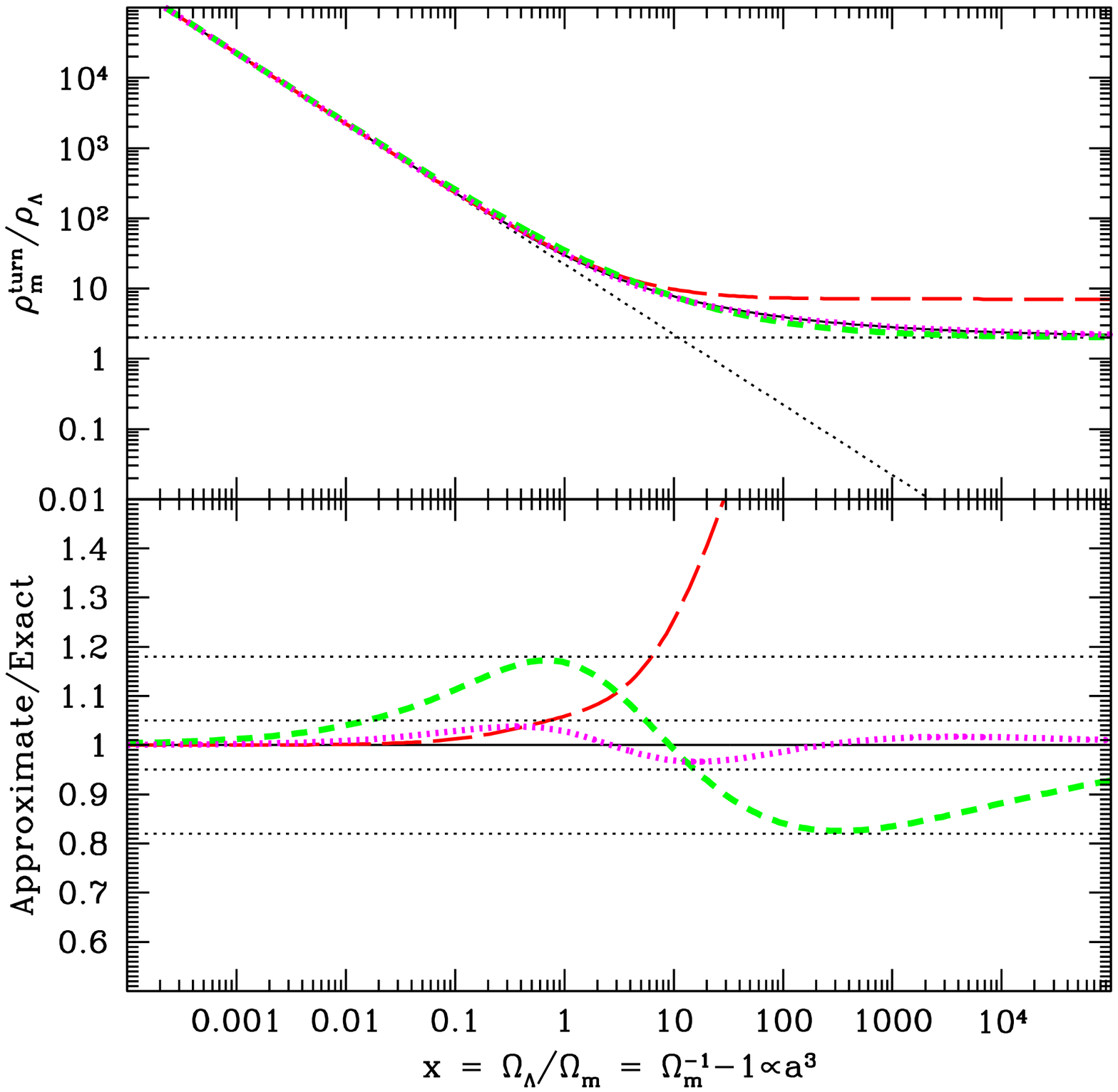}}
\caption[1]{\label{rhoturnFig}\footnotesize%
The top panel shows the turnaround (minimum) matter density $\rhoturn$ of a top hat halo as a function of the time $x$ when its collapses.
The virial density of the resulting dark matter halo is $\rhovir\sim 8\rhoturn$.
The solid curve is the exact result, with the asymptotic behavior
$\rhoturn\to 9\pi^2\rhom/4 x$ as $x\to 0$
and
$\rhoturn\to 2\rhol$ as $x\to\infty$ (dotted lines).
The three analytic fits correspond to \eq{rhoturnFitEq1} (short-dashed), 
\eq{rhoturnFitEq2} (dotted) and that of \cite{BryanNorman98,Bullock99}, which in our notation is
$(18\pi^2/x + 18\pi^2-82 - 39x/(1+x))/8$.
The first two are seen to be accurate to better than 18\% and 4\% respectively (bottom panel).
}
\end{figure}

We will now derive the virial density of a halo formed at time $x$ in a $\Lambda$CDM universe,
generalizing the standard result $\rhovir\sim 18\pi^2\rho_m(x)$.
Whereas past work on this subject \cite{BryanNorman98,Bullock99} has focused on accurate numerical fits
to the observationally relevant present epoch and recent past ($x\simlt 1$), we need  
a formula accurate for all times $x$.

Consider first the evolution of a top hat overdensity in a $\Lambda$CDM universe.
By Birkhoff's theorem, it is given by the Friedman equation corresponding to a closed universe:
\beq{FriedmanEq}
{3H^2\over 8\pi G} = \rhol + \rhom - {k\over a^2}  = (1-{\kappa\over A^2}+{1\over A^3})\rhol,
\eeq
where we have defined a dimensionless scale factor $A\equiv(\rhol/\rhom)^{1/3}$
and a dimensionless curvature parameter $\kappa\equiv (A/a)^2 k$.
One readily finds that this will give $H=0$ and recollapse
for any curvature $\kappa>\kmin\equiv{3/2^{2/3}}\sim 1.8899$, 
so the scale factor 
$\Amax(\kappa)$ at turnaround is given by solving \eq{FriedmanEq} for $H=0$,
\ie, by numerically finding the smallest positive root $A$ of the cubic polynomial $A^3-\kappa A+1=0$.
$\Amax(\kmin)=2^{-1/3}\approx 0.79$, falling off as $\Amax(\kappa)\approx 1/\kappa$ for $\kappa\gg 1$.

The age of the universe $\tturn$ when the top hat turns around is
given by integrating $dt=d\ln A/(d\ln A/dt) = dA/(AH)$ using \eq{FriedmanEq}, \ie,
\beq{TopHatAgeEq1}
H_\Lambda\tturn(\kappa) = \int_0^{\Amax(\kappa)} {dA\over\sqrt{A^2-\kappa+{1\over  A}}},
\eeq
where $H_\Lambda\equiv (8\pi G\rhol/3)^{1/2}$.
For the unperturbed background $\Lambda$CDM universe where we used $x$ as a more convenient time variable, 
we have $\kappa=0$ and $A=\xturn^{1/3}$ when the overtensity turns around, 
giving $\tturn$ in terms of $\xturn$:
\beq{TopHatAgeEq2}
H_\Lambda\tturn(\xturn) = \int_0^{\xturn^{1/3}} {dA\over\sqrt{A^2+{1\over  A}}}.
\eeq
Eliminating $\tturn$ between equations\eqn{TopHatAgeEq1}~and\eqn{TopHatAgeEq2} 
allows us to numerically compute the function $\kappa(\xturn)$.

Putting everything together now allows us to numerically compute the 
density of our top hat overdensity at turnaround: 
Since $\rhom=\rhol/A^3$, we have
\beq{rhoturnEq}
\rhoturn(\xturn) = {\rhol\over\Amax(\kappa(\xturn))^3}.
\eeq
The result is shown in \fig{rhoturnFig} together with three analytic fitting functions.
The simple fit 
\beq{rhoturnFitEq1}
\rhoturn(x)\approx 2\left[\left({9\pi^2\over 8 x}\right)^{107\over 200} + 1\right]^{200\over 107}\rhol
\eeq
is accurate to better than 18\% for all $x$ and becomes exact in the two limits $x\to 0$ and $x\to\infty$.
The fit 
\beq{rhoturnFitEq2}
\rhoturn(x)\approx \left({9\pi^2\over 4x} + {6.6\over x^{0.3}} + 2\right)\rhol 
\eeq
is accurate to better than 4\% for all $x$ and also becomes exact in both limits,
but we will nontheless use the less accurate \eq{rhoturnFitEq1} 
because it is analytically invertible and still accurate enough for our purposes.
Applying the virial theorem gives
a characteristic virial density $\rhovir\sim 8\rhoturn$, with a weak additional 
dependence on $\rhol$ \cite{Lahav91}\footnote{The detailed calculation of \cite{Lahav91} shows that
the standard collapse factor of 2 varies at thr 20\% level with $\rhol$ at the $20\%$ level.
Qualitatively, positive (repulsive) $\rhol$ increases the collapse factor because
shells have to fall further to acquire a velocity which will bring the system into equilibrium.
We will not explicitly model this correction here given the larger uncertainties 
pertaining to other aspects of our treatment.
}.
This means  that a halo that virialized at
time $x$ has
\beq{rhovirFitEq}
\rhovir\sim\left[(18\pi^2\rhom(x))^{107\over 200} + (16\rhol)^{107\over 200}\right]^{200\over 107},
\eeq
\ie, essentially the larger of the two terms $18\pi^2\rhom(x)$ and $16\rhol$.
Inverting this, we obtain
\beq{rhovirInvEq}
x \sim {9\pi^2\over 8}\left[\left({\rhovir\over 16\rhol}\right)^{107\over 200}-1\right]^{-{200\over 107}}.
\eeq

A familiar quantity in the literature is the
collapse overdensity $\Delta_c\equiv\rhovir/\rho_m=\rhovir x/\rhol$, equaling $18\pi^2\approx 178$ 
for a flat $\rhol=0$ universe.
Using \eq{rhoturnFitEq1}, \eq{rhoturnFitEq2}
and the fit in \cite{Bullock99} to the calculations of \cite{BryanNorman98}
gives
\beqa{DeltaCeq}
\Delta_c(x)&\approx&18\pi^2\left[1+\left({8x\over 9\pi^2}\right)^{107\over 200}\right]^{200\over 107},\\
\Delta_c(x)&\approx&18\pi^2 + {52.8 x^{0.7}} + 16x,\label{DeltacEq2}\\
\Delta_c(x)&\approx&18\pi^2 + (18\pi^2-82)x - 39{x^2\over 1+x},\label{DeltacEq3}
\eeqa
respectively.
\Fig{rhoturnFig} shows that our approximations substantially improve the accuracy, \eq{DeltacEq2} being good to 4\%.
This improvement not surprising, since the approximation\eqn{DeltacEq3} was not designed to work well for $x\gg 1$.

\end{document}